# Chemical control of polarization in thin strained films of a multiaxial ferroelectric: phase diagrams and polarization rotation

Anna N. Morozovska[1*], Eugene A. Eliseev[2], Arpan Biswas[3], Hanna V. Shevliakova[1], Nicholas V. Morozovsky[1], and Sergei V. Kalinin[3†]

[1] Institute of Physics, National Academy of Sciences of Ukraine, pr. Nauky 46, 03028 Kyiv, Ukraine

[2] Institute for Problems of Materials Science, National Academy of Sciences of Ukraine, Krjijanovskogo 3, 03142 Kyiv, Ukraine

[3] Center for Nanophase Materials Sciences, Oak Ridge National Laboratory, Oak Ridge, TN 37831

### Abstract

The emergent behaviors in thin films of a multiaxial ferroelectric due to an electrochemical coupling between the rotating polarization and surface ions are explored within the framework of the 2-4 Landau-Ginzburg-Devonshire (LGD) thermodynamic potential combined with the Stephenson-Highland (SH) approach. The combined LGD-SH approach allows to describe the electrochemical switching and rotation of polarization vector in the multiaxial ferroelectric film covered by surface ions with a charge density dependent to the relative partial oxygen pressure. We calculate the phase diagrams and analyze the dependence of polarization components on the applied voltage, and discuss the peculiarities of quasi-static ferroelectric, dielectric and piezoelectric hysteresis loops in thin strained multiaxial ferroelectric films. The nonlinear surface screening by oxygen ions makes the diagrams very different from the known diagrams of e.g., strained BaTiO$_3$ films. Quite unexpectedly we predict the appearance of the ferroelectric reentrant phases. Obtained results point on the possibility to control the appearance and features of ferroelectric, dielectric and piezoelectric hysteresis in multiaxial FE films covered by surface ions by varying their concentration via the partial oxygen pressure. The LGD-SH description of a multiaxial FE film can be further implemented within the Bayesian optimization framework, opening the pathway towards predictive materials optimization.

---

[*] corresponding author, e-mail: anna.n.morozovska@gmail.com
[†] corresponding author, e-mail: sergei2@ornl.gov

## I. INTRODUCTION

Since over 40 years ago, ferroelectrics (**FE**) have emerged as promising materials for electronic applications. The early proposals introduced the concepts of ferroelectric gate transistor [1, 2]. Early 90ies have seen the advances in ferroelectric random-access memories [3, 4, 5] and, spurred by the advances in piezoresponse force microscopy [6, 7, 8, 9] of ferroelectric data storage. The progress in ferroelectric film growth further led to the concepts of ferroelectric tunneling barriers and multiferroic devices in early XXI century [10, 11, 12]. Notably, many of these applications have been demonstrated as prototypes and even marketed devices. For classical oxide perovskites, the difficulties in the materials integration invariably led to the ferroelectric devices being overtaken by alternative technologies. However, the emergence of binary ferroelectrics such as hafnia and zirconia-based systems [13, 14, 15], magnesium-zinc oxides [16], and aluminum-scandium nitrides finally breaks this paradigm [17, 18], and combined with recent acute attention to microelectronic devices in US sets a stage for rapid progress in the field.

While the bulk properties of ferroelectrics are well understood and are generally amenable to the characterization by a broad variety of elastic and inelastic scattering techniques, this is not the case for ferroelectric surfaces and interfaces. In particular, the early field of ferroelectrics identified the crucial role of polarization screening on ferroelectric phase stability [19], namely that in the absence of screening or domain formation depolarization fields will suppress the ferroelectric phase [20]. At that time, it was postulated that accumulation of screening charges at the interfaces will compensate for the polarization charges, and detailed screening mechanisms were not considered. Later, the simplified approach in which potentials on surfaces are prescribed, along with the introduction of ferroelectric dead layer concept to reflect separation between screening and polarization bound charges. This dead layer model is now de-facto standard in ferroelectric field [20, 21, 22], and is used both for phenomenological modelling [23] and for the interpretation of the results of density functional modelling [24, 25, 26].

It should be noted that a rich set of work was developed in the concept of ferroelectrics-semiconductors. Here, the interplay between ferroelectric and semiconducting subsystem of the same material were considered, allowing to account for polarization dependent photovoltaic [27, 28, 29] and photoelectrochemical [30, 31] properties of ferroelectric surfaces. Interestingly, this direction now sees a resurgence with the rapid emergence of hybrid ferroelectric perovskites, in which ionic mobility is strongly coupled to photovoltaic [32, 33] and domain formation phenomena [34].

At the same time, very little theoretical work has been done in the context of ferroelectrics coupled to realistic surfaces and interfaces sustaining electrochemical and semiconductive behaviors, i.e., having finite density of electronic or ionic states [35]. These scenarios correspond to realistic boundary conditions on open ferroelectric surfaces via ionic adsorption is intrinsically coupled to the surface electrochemical processes [36, 37, 38] and in the integrated devices [23]. For open ferroelectric surfaces, there is by now the preponderant evidence of the surface screening by adsorbed ions [39, 40, 41]. Previously, the theoretical formalism for the analysis of the ferroelectric behavior in proximity to electrochemically coupled interface was developed by Stephenson and Highland (**SH**) [42, 43]. Morozovska and Kalinin group [44, 45, 46, 47] have developed it further using the Landau-Ginzburg-Devonshire-Stephenson-Highland (**LGD-SH**) approach. They derived the analytical solutions and relevant phase behaviors for the **uniaxial** ferroelectrics in 2017, and recently consider the for antiferroelectrics with electrochemical polarization switching [48, 49]. The analysis [44 – 49] leads to the elucidation of **ferroionic** states, which are the result of nonlinear electrostatic interaction between the ions with the surface charge density obeyed Langmuir adsorption isotherm and ferroelectric dipoles. The properties of these states were described by the system of coupled 1D equations. However, the influence surface ionic screening on the polar properties of **multiaxial** FE films covered by the layer of oxygen ions have not been considered theoretically, despite exactly the case seems very interesting for the fundamental research, is close to experiments and promising for applications.

To fill the gap in the knowledge, here we further extend the LGD-SH approach to the very important case of the **multiaxial** ferroelectrics, where the polarization can rotate depending on the surface electrochemical conditions and strain. Polarization rotation is broadly perceived one of the fundamental mechanisms underpinning the enhanced electromechanical and dielectric properties of disordered ferroelectrics [50, 51]. Similarly, interplay between polarization rotation enabled by flat energy landscapes and disorder is broadly considered as one of the mechanisms behind the unique functionalities of ferroelectric relaxors and morphotropic phase boundary materials [52, 53].

Below we present the formalism for this analysis, construct phase diagrams as a function of pressure, temperature, and strain, and discover highly unusual behaviors, including the reentrant phase transitions between ferroelectric, ferroionic and non-ferroelectric states. We calculate and analyze the dependence of polarization, dielectric permittivity and piezoelectric coefficients on applied voltage.

The manuscript is structured as following. **Section II** contains basic LGD equations and SH problem formulation with boundary conditions. **Section III** analyze the influence of mismatch strains and surface ions the free energy of the film. Phase diagrams, ferroelectric, dielectric and piezoelectric hysteresis loops are presented and analyzed in **Section IV**. **Section V** is a brief summary. Calculations details and auxiliary figures are listed in **Suppl. Mat**. [54].

## II LANDAU-GINZBURG-DEVONSHIRE-STEPHENSON-HIGHLAND APPROACH
### A. Problem formulation

Here we consider a system consisting of an electron-conducting substrate electrode, a multiaxial FE film of thickness $h$, covered with a layer of surface ions with a charge density $\sigma(\phi)$. An in-plane mismatch strain $u_m$ emerges due to the difference of the film and substrate lattice constants. An ultra-thin gap of width λ separates the film surface and the top electrode, that is either an ion-conductive planar electrode or a flatted apex of SPM tip. The gap provides a direct ion exchange with an ambient media, as shown in **Fig. 1(a).** In the case of perfect electric contact, $\lambda = 0$, the ion exchange is impossible.

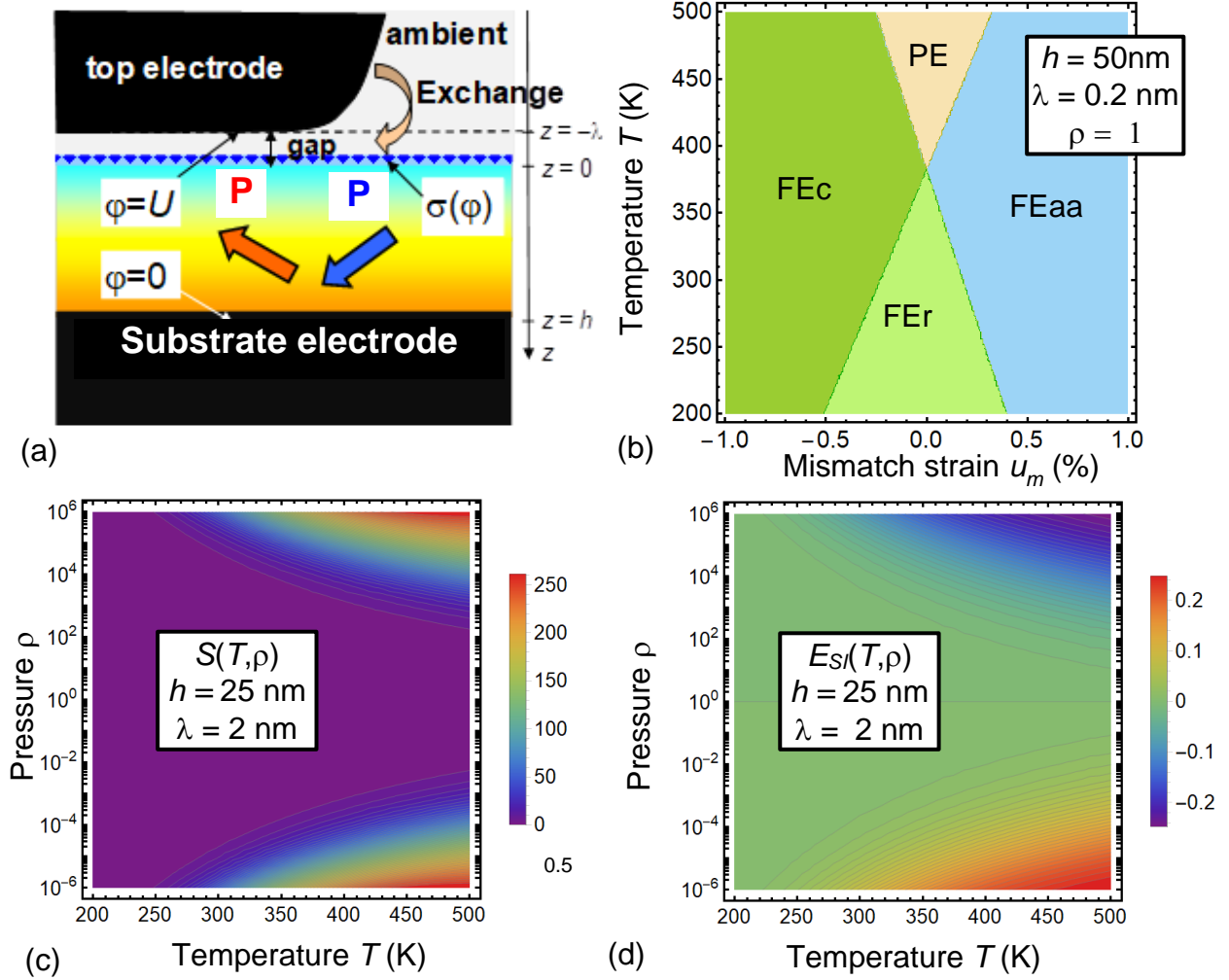

**FIGURE 1**. **(a)** Layout of the considered system, consisting of an electron-conducting substrate electrode, a multiaxial FE film of thickness $h$ with a polarization vector $\boldsymbol{P}$; a layer of surface ions with a charge density $\sigma(\phi)$, an ultra-thin gap of width $\lambda$ separating film surface, and a top electrode providing a direct ion exchange with an ambient media (from bottom to the top). Adapted from Ref. [45]. **(b)** A typical phase diagram of a multiaxial FE film covered with perfect electrodes. Color maps of the function $S(T, \rho, h)$ **(c)** and built-in field $E_{SI}(T, \rho, h)$ **(d)** in dependence on temperature $T$ and relative oxygen pressure $\rho$ calculated for $h = 25$ nm and $\lambda = 2$ nm. Other parameters and designations are listed in **Tables I-III.**

Due to the presence of an ultra-thin dielectric gap between the top electrode and the surface of the film, the linear equation of state $\boldsymbol{D} = \varepsilon_0 \varepsilon_d \boldsymbol{E}$ relates an electric displacement $\boldsymbol{D}$ and field $\boldsymbol{E}$ in the gap. Here $\varepsilon_0$ is a universal dielectric constant and $\varepsilon_d \sim (1 - 10)$ is a relative permittivity in the gap filled by an air with controllable oxygen pressure. A wide band-gap FE film can be considered insulating, and here $\boldsymbol{D} = \varepsilon_0 \boldsymbol{E} + \boldsymbol{P}$. A potential $\phi$ of a quasi-static electric field satisfies a Laplace equation in the gap and a Poisson equation in the film. The boundary conditions for the system are the equivalence of the electric potential to the applied voltage $U$ at the top electrode (modeled by a flattened region $z = -\lambda$); and the equivalence of the difference $D_3^{(gap)} - D_3^{(film)}$ to the ionic surface charge density $\sigma[\phi(\vec{r})]$ at $z = 0$; the continuity of the $\phi$ at gap - film interface $z = 0$; and zero potential at the conducting bottom electrode $z = h$ [see **Fig. 1(a)**].

The polarization components of the multiaxial FE film depend on the electric field $E_i$ as $P_i = P_i^f + \varepsilon_0(\varepsilon_{ij}^b - 1)E_j$, where $i = 1,2,3$, and $\varepsilon_{33}^b$ is a relative background permittivity of antiferroelectric, $\varepsilon_{ij}^b \leq 10$ [55]. The polarization component $P_i^f$ is further abbreviated as $P_i$. To determine the spatial-temporal evolution of $P_i$, we use the Landau-Ginzburg-Devonshire (LGD) approach, as explained in **Appendix A** [54].

The Landau-Ginzburg-Devonshire (**LGD**) free energy functional $F$ additively includes a bulk part – an expansion on 2-4-6 powers of the polarization components $P_i$, $F_{bulk}$; a polarization gradient energy contribution, $F_{grad}$; an electrostatic and electrostriction contributions, $F_{el}$; and a surface energy, $F_S$:

$$F = F_{bulk} + F_{grad} + F_{el} + F_S, \tag{1}$$

The contributions to Eq.(1) are

$$F_{bulk} = \int_{V_f} d^3r \left( a_i P_i^2 + a_{ij} P_i^2 P_j^2 + a_{ijk} P_i^2 P_j^2 P_k^2 \right), \tag{2a}$$

$$F_{grad} = \int_{V_f} d^3r \frac{g_{ijkl}}{2} \left( \frac{\partial P_i}{\partial x_j} \frac{\partial P_k}{\partial x_l} \right), \tag{2b}$$

$$F_{el} = -\int_{V_f} d^3r \left( P_i E_i + \frac{\varepsilon_0 \varepsilon_b}{2} E_i E_i + Q_{ijkl} P_i P_j u_{kl} \right), \tag{2c}$$

$$F_S = \frac{1}{2} \int_S d^2r \, a_{ij}^{(S)} P_i P_j. \tag{2d}$$

Here $V_f$ is the film volume. The coefficients $a_i$ linearly depends on the temperature $T$. In general, the tensor $a_{ijk}$ and the polarization gradient tensor $g_{ijkl}$ are positively defined and also regarded temperature-independent. Hereinafter we regard that $a_{ij}$ is positively defined and then one can disregard $a_{ijk}$ for the second order ferroelectrics.

The influence of elastic strain $u_{kl}$ (e.g., mismatch strain) via electrostriction coupling (i.e., via the term $Q_{ijkl} P_i P_j u_{kl}$) can be approximately taken into account by the renormalization of coefficients $a_{ij}$ and $a_{ijk}$ in Eq.(2a) (see e.g., [56]). The electrostatic energy is dependent on the surface ionic charge $\sigma(\phi)$, since the electric field is charge-dependent.

An equation for the surface ionic charge density $\sigma(\phi)$ is analogous to the Langmuir adsorption isotherm used in interfacial electrochemistry for adsorption onto a conducting electrode exposed to ions in a solution [57]. To describe the dynamics of the surface charge density, we use a linear relaxation model,

$$\tau \frac{\partial \sigma}{\partial t} + \sigma = \sigma_0[\phi], \tag{3}$$

where the dependence of an equilibrium charge density $\sigma_0[\phi]$ on the electric potential $\phi$ is controlled by the concentration of surface ions, $\theta_i(\phi)$, at the interface $z = 0$ in a self-consistent manner, as proposed by Stephenson and Highland [43]:

$$\sigma_0[\phi] = \sum_{i=1}^{2} \frac{eZ_i \theta_i(\phi)}{A_i} \equiv \sum_{i=1}^{2} \frac{eZ_i}{A_i} \left(1 + \rho^{1/n_i} \exp\left(\frac{\Delta G_i^{00} + eZ_i\phi}{k_B T}\right)\right)^{-1}. \tag{4}$$

Here $e$ is an elementary charge, $Z_i$ is the ionization degree of the surface ions/electrons, $1/A_i$ are saturation densities of the surface ions. A subscript $i$ designates the summation on positive ($i = 1$) and negative ($i = 2$) charges, respectively; $\rho = \frac{p_{O2}}{p_{O2}^{00}}$ is the relative partial pressure of oxygen (or other ambient gas) [43], $n_i$ is the number of surface ions created per gas molecule. Two surface charge species exist, since the gas molecule had been electroneutral before its electrochemical decomposition started. The dimensionless ratio $\rho$ varies in a relatively wide range from $10^{-6}$ to $10^6$ [42, 43].

Positive parameters $\Delta G_1^{00}$ and $\Delta G_2^{00}$ are the free energies of the surface defects formation at normal conditions, $p_{O2}^{00} = 1$ bar, and zero applied voltage $U = 0$. The energies $\Delta G_i^{00}$ are responsible for the formation of different surface charge states (ions, vacancies, or their complexes). Specifically, exact values of $\Delta G_i^{00}$ are poorly known even for many practically important cases, and so hereinafter they are regarded varying in the range ~(0 − 1) eV [42, 43]. Notably, the developed solutions are insensitive to the specific details of the charge compensation process [58], but are sensitive to the thermodynamic parameters of corresponding reactions [59].

It is important to note that while Eq.(4) pertains to a specific case of surface electrochemical reaction, the proposed formalism is considerably more universal. The equations (1)-(4) jointly define the coupling between the surface potential and the potential-dependent interface screening charge. As such, this formalism can be readily adapted to other electrochemical models, or to the specific functional form of interface density of electronic or ionic states.

## B. Single-domain approximation

Since the stabilization of a single-domain polarization in ultrathin perovskite films takes place due to the chemical switching (see e.g. Refs.[39 - 43]), we will assume that the distribution $P_i(x, y, z)$ does not deviate significantly from the values $\langle P_i \rangle$ averaged over the film thickness, namely $P_i \cong \langle P_i \rangle$. Neglecting the polarization gradient energy, the bulk density of the potential (2) acquires the form:

$$f_{bulk} = a_i P_i^2 + a_{ij} P_i^2 P_j^2 + a_{ijk} P_i^2 P_j^2 P_k^2 + Q_{ijkl} P_i P_j u_{kl} - P_3 E_3. \tag{5}$$

Here we assume that the in-plane components of electric field are absent, $E_1 = E_2 = 0$, and consider the out-of-plane electric field component $E_3$ only. It consists of the external field, depolarization field and the field created by the surface ion layer. Expression for component $E_3$ follows from the minimization of electrostatic energy, written in the form [48]:

$$f_{el} = -\Psi P_3 - \varepsilon_0 \varepsilon_{33}^b \frac{\Psi^2}{2h} - \frac{\varepsilon_0 \varepsilon_d}{2} \frac{(\Psi - U)^2}{\lambda} + \int_0^\Psi \sigma_0[\varphi] d\varphi. \tag{6a}$$

The field $E_3$ has the form:

$$E_3 = \frac{\Psi}{h} = \frac{\lambda(\sigma_0[\Psi] - P_3) + \varepsilon_0 \varepsilon_d U}{\varepsilon_0(\varepsilon_d h + \lambda \varepsilon_{33}^b)}. \tag{6b}$$

Eq.(6b) corresponds either to the stationary case, or to the adiabatic conditions, when the surface charge density $\sigma = \sigma_0$, and the dependence $\sigma_0[\Psi]$ is given by Eq.(4). The overpotential $\Psi$ contains the contribution from surface charges proportional to $\sigma_0$, the depolarization field contribution proportional to $P_3$, and the external potential drop proportional to applied voltage $U$.

## III. FREE ENERGY OF ANTIFERRO-IONIC SYSTEM. POSSIBLE PHASES
### A. The influence of mismatch strains and surface ion charge on free energy coefficients

Following Pertsev et.al. [56], the coefficients $a_i$ in Eq.(5a) must be renormalized by a mismatch strain $u_m$, which can change their sign for high enough compressive or tensile $|u_m|$. The coefficients $a_{ij}$ in the same equation can change sign from negative to positive due to the electrostriction coupling with a substrate. Below we consider the FE films with a cubic m3m symmetry of the parent phase that allows us to use analytical results [56] for the renormalization of the coefficients $a_i$ and $a_{ij}$ by the mismatch strain and electrostriction.

Assuming that $|eZ_i\Psi/k_BT| \ll 1$ the potential density (5) can be further expanded in $P_i$ and $\Psi$ powers [47 - 49]. In result we obtain the expression for the "effective" free energy from Eq.(5):

$$f_R = b_1(P_1^2 + P_2^2) + b_3P_3^2 + b_{11}(P_1^4 + P_2^4) + b_{33}P_3^4 + b_{12}P_1^2P_2^2 + b_{13}(P_1^2 + P_2^2)P_3^2 - P_3\tilde{E}_3, \quad (7)$$

which coefficients derived in **Appendix B** [54] depend on the mismatch strain $u_m$, oxygen pressure $\rho$ and temperature $T$ in the following way:

$$b_3(T,\rho,h,u_m) = \left(a_1(T) - \frac{2Q_{12}u_m}{s_{11}+s_{12}}\right)(1 + S(T,\rho,h)) + \frac{\lambda}{2\varepsilon_0(\varepsilon_d h + \lambda\varepsilon_{33}^b)}, \quad (8a)$$

$$b_1(T, u_m) = a_1(T) - u_m\frac{Q_{11}+Q_{12}}{s_{11}+s_{12}}, \quad (8b)$$

$$b_{11} = a_{11} + \frac{s_{11}(Q_{11}^2+Q_{12}^2) - 2Q_{11}Q_{12}s_{12}}{2(s_{11}^2 - s_{12}^2)}, \quad (8c)$$

$$b_{33} = \left(a_{11} + \frac{Q_{12}^2}{s_{11}+s_{12}}\right)(1 + S(T,\rho,h)), \quad (8d)$$

$$b_{12} = a_{12} - \frac{s_{12}(Q_{11}^2+Q_{12}^2) - 2Q_{11}Q_{12}s_{11}}{s_{11}^2 - s_{12}^2} + \frac{Q_{44}^2}{2s_{44}}, \quad (8e)$$

$$b_{13}(T,\rho,h) = \left(a_{12} + \frac{Q_{12}(Q_{11}+Q_{12})}{s_{11}+s_{12}}\right)(1 + S(T,\rho,h)). \quad (8f)$$

The coefficient $a_1(T) = \alpha_T(T - T_P)$ changes the sign at Curie temperature $T_P$, and the inverse Curie constant $\alpha_T > 0$. Coefficients $Q_{ij}$ are electrostriction tensor components and $s_{ij}$ are elastic compliances. The inequalities $b_{11} > 0$, $b_{33} > 0$, $b_{12} > -2b_{11}$ and $b_{13} > -2\sqrt{b_{11}b_{33}}$ should be valid for the LGD potential stability. The first term in Eq.(8a), proportional to $(1 + S(T,\rho,h))$, is renormalized by the influence of the surface charge caused by the adsorption/desorption of the oxygen ions via the function $S(T,\rho,h)$. The second term in Eq.(8a) proportional to the gap width, $\frac{\lambda}{2\varepsilon_0(\varepsilon_d h + \lambda\varepsilon_{33}^b)}$, originates from the depolarization field. Also, we introduce the following positive functions in Eqs.(8a) and (8f):

$$S(T,\rho,h) = \frac{\lambda h}{\varepsilon_0(\varepsilon_d h + \lambda\varepsilon_{33}^b)}\sum_{i=1,2}\frac{(eZ_if_i(T,\rho))^2}{A_ik_BT}, \quad f_i(T,\rho) = \left(1 + \rho^{1/n_i}\exp\left(\frac{\Delta G_i^{00}}{k_BT}\right)\right)^{-1}. \quad (9)$$

The effective electric field $\tilde{E}_3$ in Eq.(7) is a sum of a built-in field $E_{SI}$ and an acting field $E_a$:

$$\tilde{E}_3(T,\rho,h) = E_{SI}(T,\rho,h) + E_a(U,h), \quad (10a)$$

$$E_{SI}(T,\rho,h) = \frac{\lambda}{\varepsilon_0(\varepsilon_d h + \lambda\varepsilon_{33}^b)}\sum_{i=1,2}\frac{eZ_i}{A_i}f_i(T,\rho), \quad (10b)$$

$$E_a(U,h) = -\frac{\varepsilon_d U}{\varepsilon_d h + \lambda \varepsilon_{33}^b}. \tag{10c}$$

The built-in field $E_{SI}(T,\rho,h)$ created by surface ions is significant for thin film, being proportional to the ratio $\frac{\lambda}{h}$. Since, as a rule, $\varepsilon_d h \gg \lambda \varepsilon_{33}^b$, the acting field is close to an external field, $E_a \approx E_e = -\frac{U}{h}$. So, we need to play not only with the oxygen pressure $\rho$ and temperature $T$, but also need support positive $b_{11}$, $b_{12}$ and $b_{33}$ in the actual temperature range by the electrostriction coupling. We also need to select an appropriate mismatch strain to support either in-plane or out-of-plane polarization that couples with a surface electrochemistry.

It is seen from Eqs.(8)-(10) that the influence of the relative partial oxygen pressure on the multiaxial FE film reduces to the influence of the surface charge $\sigma_0[\phi]$ caused by the adsorption/desorption of oxygen ions, mathematically expressed via the positive function $S(T,\rho,h)$ and the built-in electric field $E_{SI}(T,\rho,h)$ penetrating the entire the depth of the film.

The behavior of the polarization components, $P_i$, can be described via relaxation-type nonlinear differential equations, $\frac{\partial f_R}{\partial P_i} = -\Gamma \frac{\partial P_i}{\partial t}$, similar to the ones derived in Refs. [44]:

$$\Gamma \frac{\partial P_1}{\partial t} + (2b_1 + 4b_{11}P_1^2 + 2b_{12}P_2^2 + 2b_{13}P_3^2)P_1 = 0, \tag{11a}$$

$$\Gamma \frac{\partial P_2}{\partial t} + (2b_1 + 4b_{11}P_2^2 + 2b_{12}P_1^2 + 2b_{13}P_3^2)P_2 = 0, \tag{11b}$$

$$\Gamma \frac{\partial P_3}{\partial t} + \left(2b_3 + 4b_{33}P_3^2 + 2b_{13}(P_1^2 + P_2^2)\right)P_3 = E_{SI}(T,\rho,h) + E_a(U,h). \tag{11c}$$

Eqs.(11) follow from the variation of the free energy (7). Here $\Gamma_{P,A}$ are the positive kinetic coefficients defining the Khalatnikov relaxation of the order parameters.

### B. The table of homogeneous phases and material parameters

Homogeneous FE phases, the spontaneous order parameters and free energy (5) of the phases at $E_i = 0$ are listed in **Table I**. The abbreviations "PE" and "FE" means the paraelectric and ferroelectric phases, respectively. The classification and designations of FE phases are very-well known [56] and come from the classification of domains with different orientation of polarization. They are based on the letter-type "code" for zero and nonzero components of **P** under the absence of total electric field. The nonzero components 1, 2 and 3 are denoted with letters "a", "b" or "c", respectively. If the same letter "a" is used twice, it means that $P_1 = \pm P_2$. For the sake of the brevity, we use the letter "r" for the "abc" or "aac" cases, since it is a rhombohedral phase.

We also note that the built-in electric field $E_{SI}$ cannot induce new FE phases with reversible polarization, but it can induce the electret-like ferrielectric (**FEI**) phase (not listed in **Table I**) instead of PE phase (listed in **Table I**). The FEI phase is characterized by a spontaneous polarization that is absent in the PE phase. At that the boundary between the FEI with PE phase is diffuse, and only those part of PE phase corresponding to $\rho \ll 1$ (or $\rho \gg 1$) becomes FEI phase. Of course, the built-in field can induce the polarization component $P_3$ in the FEaa phase, but it is electret-like and not bistable. Simple analytical expressions for the FEI phase energy and electret-like polarization components are absent.

**Table I.** Homogeneous ferroelectric phases, the spontaneous order parameters and free energy of the phases at zero total field, $E_{SI}(T, \rho, h) + E_a(U, h) = 0$.

| Phase | Spontaneous order parameters | Free energy $f_R$ |
|---|---|---|
| PE | $P_1 = P_2 = P_3 = 0,$ | 0 |
| FEa | $P_1 = \pm\sqrt{-\frac{b_1}{2b_{11}}}, P_2 = P_3 = 0$ | $-\frac{b_1^2}{4b_{11}}$ |
| FEc | $P_3 = \pm\sqrt{-\frac{b_3}{2b_{33}}}, P_1 = P_2 = 0$ | $-\frac{b_3^2}{4b_{33}}$ |
| FEaa | $P_1 = \mp P_2 = \pm\sqrt{-\frac{b_1}{2b_{11}+b_{12}}}, P_3 = 0$ | $-\frac{b_1^2}{(2b_{11}+b_{12})}$ |
| FEac | $P_1 = \pm\sqrt{-\frac{2b_{33}b_1 - b_{13}b_3}{4b_{11}b_{33} - b_{13}^2}}, P_3 = \pm\sqrt{-\frac{2b_{11}b_3 - b_{13}b_1}{4b_{11}b_{33} - b_{13}^2}}, P_2 = 0$ | $\frac{-b_{33}b_1^2 - b_{11}b_3^2 + b_{13}b_1 b_3}{(4b_{11}b_{33} - b_{13}^2)}$ |
| FEr | $P_1 = \mp P_2 = \frac{\pm\sqrt{b_3 b_{13} - 2 b_1 b_{33}}}{\sqrt{2(2b_{11}+b_{12})b_{33} - 2b_{13}^2}}, P_3 = \pm\frac{\sqrt{-b_3(2b_{11}+b_{12}) + 2 b_1 b_{13}}}{\sqrt{2(2b_{11}+b_{12})b_{33} - 2b_{13}^2}}$ | $\frac{-b_3^2(2b_{11}+b_{12}) - 4b_1^2 b_{33} + 4 b_1 b_3 b_{13}}{4\left((2b_{11}+b_{12})b_{33} - b_{13}^2\right)}$ |

**Table II** contains the material parameters of a hypothetic bulk ferroelectric close to BaTiO$_3$, used in this work. From the table, the coefficient $a_{11} = 3.6\,(T+222)\cdot 10^6$ m$^5$J/C$^4$ is positive at all temperatures. In contrast, the value $a_{11} = 3.6(T - 448)\cdot 10^6$ m$^5$J/C$^4$, used by Pertsev et al [56], is negative below 448 K. The electrostriction coupling term, $\frac{s_{11}(Q_{11}^2 + Q_{12}^2) - 2 Q_{11} Q_{12} s_{12}}{2(s_{11}^2 - s_{12}^2)}$, appeared too small to make the Pertsev et al. coefficient $b_{11}$ positive above 245 K, requiring inclusion of the higher terms in Eq.(5), which are proportional to $a_{ijk}$. Since the higher terms inclusion does not allow us to derive analytical results, we use positive $a_{11}$ from **Table II** and neglect the higher terms. The results listed in **Table I** are used in this work, and, as it will be demonstrated in next section, the selected positive value of $a_{11}$ leads to the phase diagram structure similar to the one calculated by Pertsev et al. for BaTiO$_3$ (compare **Fig. 1(b)** and Fig. 1(a) from Ref. [56]).

**Table II.** LGD coefficients and other material parameters of a bulk BaTiO$_3$

| Coefficient | Numerical value | Refs |
|---|---|---|
| $\varepsilon_{33}^b$ (dimensionless) | $\varepsilon_{33}^b = 7$ (core background) | 60 |
| $a_i$ (mJ/C$^2$) | $a_1 = 3.3(T-383)\cdot 10^5$, $\alpha_T = 3.3\cdot 10^5$ | 56 |
| $a_{ij}$ (m$^5$J/C$^4$) | $a_{11} = 3.6(T+222)\cdot 10^6$ (this work), $a_{12} = 4.9\cdot 10^8$ (ref. 56) | this work |
| $Q_{ij}$ (m$^4$/C$^2$) | $Q_{11}=0.11$, $Q_{12}= -0.043$, $Q_{44}=0.059$ | 61 |
| $s_{ij}$ (Pa$^{-1}$) | $s_{11}=8.3\cdot 10^{-12}$, $s_{12}= -2.7\cdot 10^{-12}$, $s_{44}=9.24\cdot 10^{-12}$ | 62 |
| $g_{ij}$ (m$^3$J/C$^2$) | $g_{11}=5.0\cdot 10^{-10}$, $g_{12}= -0.2\cdot 10^{-10}$, $g_{44}= 0.2\cdot 10^{-10}$ | 60 |
| $\Gamma$ (s× C$^{-2}$ J m) | $10^{-11} - 10^{-13}$ (rather small far from $T_C$) | 60 |
| $h$ (nm) | 5 – 50 | variable |

**Table III** contains surface ion-related, geometrical and interfacial parameters. As follows from the table, the function

$$S(T,\rho,h) = S\left(T,\frac{1}{\rho},h\right) \sim \frac{(2e)^2}{Ak_BT}\left(\frac{1}{1+\rho^2 \exp\left(\frac{\Delta G}{k_BT}\right)} + \frac{1}{1+\rho^{-2}\exp\left(\frac{\Delta G}{k_BT}\right)}\right)^{-2} \quad (12a)$$

in accordance with Eq.(9), because $n_1 = -n_2 = 2$, $Z_1 = -Z_2 = 2$, $A_1 = A_2 = A$ and $\Delta G_1^{00} = \Delta G_2^{00} = \Delta G$. Thus, the physical state calculated for $\rho \neq 1$ coincides with the state calculated for $1/\rho$. So that hereinafter we analyze the case $0 \leq \rho \leq 1$ only. The condition $0 \leq \rho \leq 1$ is easier realizable experimentally in comparison with the condition $0 \leq \rho < \infty$. We also note that the built-in field

$$E_{SI}(T,\rho,h) = -E_{SI}\left(T,\frac{1}{\rho},h\right) \sim \frac{2e}{A}\left(\frac{1}{1+\rho^2 \exp\left(\frac{\Delta G}{k_BT}\right)} - \frac{1}{1+\rho^{-2}\exp\left(\frac{\Delta G}{k_BT}\right)}\right) \quad (12b)$$

in accordance with Eq.(10b); so it vanishes at $\rho \to 1$ and changes its sign under the condition $\rho \to \frac{1}{\rho}$ [see **Fig. 1(d)** for details].

**Table III.** Geometrical, surface ion-related and interfacial parameters

| Description | Designation and dimensionality | Substrate/ BaTiO$_3$ film / ionic charge / gap / tip |
|---|---|---|
| Film thickness | $h$ (nm) | vary from 5 to 50 nm |
| Mismatch strain | $u_m$ (dimensionless, in %) | variable |
| Surface charge density | $\sigma(\phi,t)$ (C/m$^2$) | variable |

| Equilibrium surface charge density | $\sigma_0(\phi)$ (C/m$^2$) | variable |
|---|---|---|
| Occupation degree of surface ions | $\theta_i$ (dimensionless) | variable |
| Relative oxygen partial pressure | $\rho$ (dimensionless) | Vary from $10^{-6}$ to $10^6$ |
| Surface charge relaxation time | $\tau$ (s) | $\gg$ Landau-Khalatnikov time |
| Width of the dielectric gap | $\lambda$ (nm) | 0.2 - 2 |
| Permittivity of the dielectric gap | $\varepsilon_d$ (dimensionless) | 1 - 10 |
| Universal dielectric constant | $\varepsilon_0$ (F/m) | 8.85×10$^{-12}$ |
| Electron charge | $e$ (C) | 1.6×10$^{-19}$ |
| Ionization degree of the surface ions | $Z_i$ (dimensionless) | $Z_1 = +2$, $Z_2 = -2$ |
| Number of surface ions created per oxygen molecule | $n_i$ (dimensionless) | $n_1 = 2$, $n_2 = -2$ |
| Saturation area of the surface ions | $A_i$ (m$^2$) | $A_1 = A_2 = A = 10^{-18}$ |
| Surface defect/ion formation energy | $\Delta G_i^{00}$ (eV) | 0.1 - 0.3, we suggest $\Delta G_1^{00} = \Delta G_2^{00} = \Delta G = 0.2$ |

In accordance with **Table III**, $\frac{\Delta G_i^{00}}{k_B T} > 4$ for $\Delta G_i^{00} \geq 0.1$ eV and $|T| < 500$ K, making the factor $\exp\left(\frac{\Delta G_i^{00}}{k_B T}\right) \gg 1$ in the actual temperature range. Therefore, the function $S(T, \rho, h)$ can strongly changes with pressure. Actually, $S(T, \rho, h)$ is minimal for $\rho = 1$, namely $S(T, 1, h) \ll 1$ at $T = 200$ K. At the same time $S(T, \rho, h)$ increases monotonically with $\rho$ deviation from unity reaching several hundred at $\rho = 10^{-6}$ and $T = 500$ K [see **Fig. 1(c)** for details]. Hence the multiplier $(1 + S(T, \rho, h))$ can significantly increase the absolute value of the negative coefficient $\left(a_1(T) - \frac{2Q_{12}u_m}{s_{11}+s_{12}}\right)$. In result the coefficient $b_3(T, \rho, h, u_m)$ can become more negative than $b_1(T, u_m)$. The circumstance opens the possibility of the out-of-plane polarization emergence at high ($\rho \gg 1$) or low ($\rho \ll 1$) partial oxygen pressures in ultra-thin multiaxial FE films. Below we will demonstrate the possibility and analyze the conditions of its appearance.

## IV. RESULTS

### A. Phase diagrams in coordinates temperature-mismatch strain

The temperature-mismatch phase diagram calculated for a thick multiaxial FE film covered with perfect electrodes is shown in **Fig. 1(b)**. The film is short-circuited, i.e., $E_a = 0$. Since the gap and oxygen pressure effect is absent for the film, the diagram is very similar to the diagram of a strained BaTiO$_3$ film calculated by Pertsev et al [56]. The values of the used constants are listed in **Table II**. The phase diagrams are calculated using the formulae in **Table I** and visualized in Mathematica 12.2 [63].

There are PE phase and three FE phases in **Fig. 1(b)**, namely the FEc phase with nonzero out-of-plane polarization $P_3$, the FEaa phase with nonzero in-plane polarization components, $P_1 = P_2$, and mixed FEr phase with nonzero $P_3$ and $P_1 = P_2$. The FEc phase is stable at compressive mismatch strains $u_m < 0$ (mainly $u_m < -0.5\%$), the PE is stable at the temperatures higher than 400 K and small $|u_m| < 0.5\%$, the FEaa phase is stable at tensile strains $u_m > 0$ (mainly $u_m > -0.5\%$), and FEr phase is stable at $T < 400$ K and $|u_m| < 0.5\%$. The boundaries between these four phases are straight lines, which cross in a single point, where all the phases have the same energy. In accordance with **Table I**, the PE-FEaa boundary is given by equation $b_1(T, u_m) = 0$, the PE-FEc boundary is given by equation $b_3(T, u_m) = 0$, FEc-FEr and FEr-FEaa boundaries are given by the equality of corresponding energies.

For the prototype BaTiO3 parameters as listed in **Table II**, compressive strains support the out-of-plane polarization component $P_3$, since $Q_{12} < 0$ and the negative term $-\frac{2Q_{12}u_m}{s_{11}+s_{12}}$ decreases the coefficient $b_3$ in Eq.(8a). Tensile strains support the in-plane polarization components $P_{1,2}$, since $Q_{11} + Q_{12} > 0$ and the negative term $-u_m \frac{Q_{11}+Q_{12}}{s_{11}+s_{12}}$ decreases the coefficient $b_1$ in Eq.(8b).

The diagrams in **Fig. 2** illustrate how the presence of oxygen pressure (we put here $\rho = 10^{-6}$) in the gap of width $\lambda = (0.4 - 2)$ nm can change the diagram shown in **Fig. 1(b)**. The diagrams in **Fig. S1-S3**, which are calculated for several pressures $\rho = 1, 10^{-2}, 10^{-4}, 10^{-6}$, gap width $\lambda = 0.2, 0.4, 2$ nm and film thickness $h =$ 50, 25 and 10 nm, illustrate the trends shown in **Fig. 2** in more details.

The diagrams in **Figs. S1-S3**, calculated for the case of zero built-in field, $E_{SI} = 0$, contain the PE phase, shown by a russet color. The built-in field vanishes at $\lambda \to 0$ or/and $\rho \to 1$ [see Eq.(10b) for details]. When the value of $E_{SI}$ increases (either with $\rho$ deviation from unity or with $\lambda$ increase) the PE phase continuously transforms in the electret-like FEI phase, which ferroionic features is especially pronounced for thin films ($h < 20$ nm) and wide gaps ($\lambda > 1$ nm). Since $E_{SI}$ gradually changes with pressure $\rho$ and temperature $T$, the boundary between the FEI and PE phases is diffuse [e.g., the remainder of PE is shown by a russet color in **Fig. 2(c)**]. The classification of other FE phases, listed in **Table I**, remains valid for $E_{SI} \neq 0$, with the remark that FEaa phase includes the hysteresis-less component $P_3$ proportional to the acting field $E_{SI} + E_a$. Note that the diagrams plotted for $\rho = 10^{-6}$ coincides with the ones for $\rho = 10^6$, that is evident from Eq.(4) and (9) for the parameters listed in **Table III**.

The most important differences between the diagram **1(b)** and **2(a)**, that is calculated for a relatively thick film ($h = 50$ nm) and narrow gap ($\lambda = 0.4$ nm), are the strongly curved boundary between FEr and FEaa phases, slightly curved boundary between FEc and FEr phases, and, not less important, the wider region of FEaa phase in **Fig. 1(b)**. The curved phase boundaries reflect the nonlinear nature of SH model for the surface charge; and the curvature increases strongly with the film thickness decrease [compare **Figs. 2(a)** for $h = 50$ nm with **Fig. 2(b)** for $h = 25$ nm]. For ultra-thin films the region of FEr phase stability becomes small and acquires the shape of rounded triangle, and the FEc phase becomes "reentrant", since the second region of the phase appear at $250 \text{ K} < T < 350 \text{ K}$ and $u_m < -0.5\%$ [see **Fig.2(c)** for $h = 10$ nm]. The reentrant FEc phase appears for wider gaps ($\lambda = 2$ nm) and borders with FEI and FEr phases, at that one of the boundaries is strongly curved (see **Fig.2(d)** for $h = 50$ nm). The area of reentrant FEc phase very weakly decreases for thinner films with a wider gap (compare **Fig.2(e)** and **2(f)** for $h = 25$ nm and $h = 10$ nm, respectively).

To resume, the diagrams shown in **Figs.2(c-f)** look very different from the diagram shown in **Fig.1(b)**, and the most interesting feature is the appearance of FEc reentrant phases. For wide gaps the reentrant FEc phase region has an unusual parabolic-like shape, shown in **Figs.2(c-f)**. The reentrance originates from the surface screening by oxygen ions. Mathematically, the features are defined by the dependence of the coefficient $b_3$ on the variables $T$ and $\rho$ (see Eqs.(8a) for details). The coefficient contains the strongly varying factor $1 + S(T, \rho, h)$. The temperature-pressure color maps of the function $S(T, \rho, h)$ are shown in **Figs.1(c)-(d)**. The behavior of $S(T, \rho, h)$ defines the nonmonotonic dependence of the coefficient $b_3$ on $T$. In results $b_3(T, \rho, h, u_m)$ can change its sign several times for compressive strains $u_m < 0$ if $\rho \ll 1$ (or $\rho \gg 1$). In accordance with **Table I**, the condition $b_3(T, \rho, h, u_m) = 0$ gives us the boundaries between PE and FEc phases. However, it appeared that the same condition corresponds to the boundary between FEr and FEaa phases with a very high accuracy [an example of such boundary is shown by a red line in **Fig. 2(f)**]. The explanation of the result is the following. Per **Table I**, the boundary between FEr and FEaa phases is given by the equivalence of their energies, $-\frac{b_1^2}{(2b_{11}+b_{12})} = \frac{-b_3^2(2b_{11}+b_{12})-4b_1^2 b_{33}+4b_1 b_3 b_{13}}{4\left((2b_{11}+b_{12})b_{33}-b_{13}^2\right)}$. Putting $b_3 = 0$ in the right-hand side, we obtain the expression $\frac{-b_1^2}{(2b_{11}+b_{12})-\frac{b_{13}^2}{b_{33}}}$ that is almost equal to the left-hand side, since the strong inequality, $\frac{b_{13}^2}{b_{33}} \ll 2b_{11} + b_{12}$, is valid.

Hence the differences between **Figs. 2(d-f)** and **Fig. 1(b)** evidently originate from the surface ions, which influence is controlled by the parameter $\rho$ and finite size effects, whose strength is controlled by the film thicknesses $h$ and gap width $\lambda$, respectively. The reentrant phase and other features are observed for compressed films (e.g., for $u_m < -0.2\%$), since the compressive strain supports the out-of-plane polarization component $P_3$ [56], and only $P_3$ is sensitive to the screening provided by the surface ions. Hence, we focus on the analysis of the influence of $\rho$, $h$ and $\lambda$ on the phase diagrams, polar and dielectric properties of the compressed multiaxial ferroelectric films.

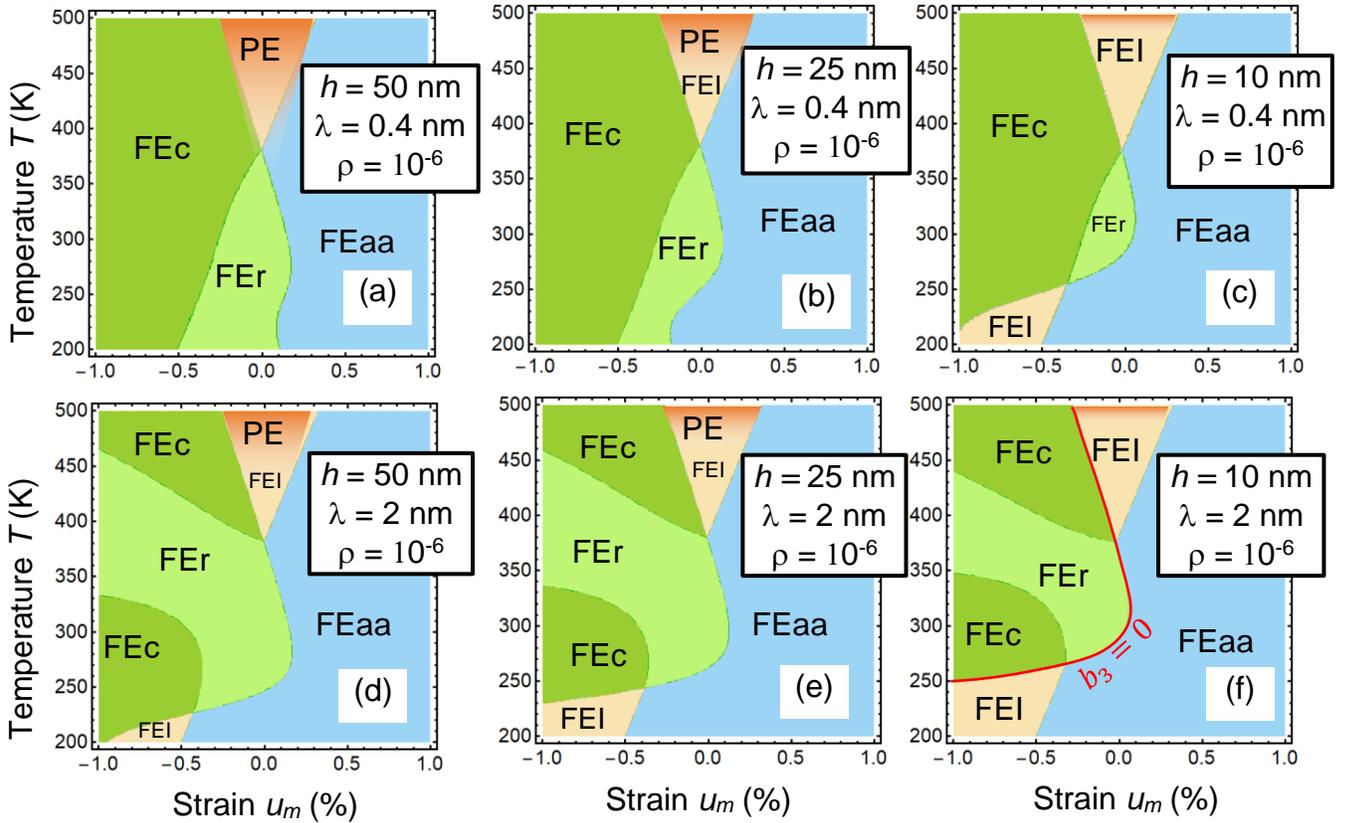

**FIGURE 2.** Typical phase diagrams of a multiaxial FE film in dependence on the temperature $T$ and mismatch strain $u_m$. A red curve in the plot (f) corresponds to the condition $b_3 = 0$. The values of relative partial oxygen pressure $\rho$, gap thickness $\lambda$ and film thickness $h$ are listed for each plot. Other parameters are listed in **Tables II-III.**

## C. Chemical control of extrinsic size effect

Typical phase diagrams of a compressed multiaxial FE film in dependence on the temperature $T$, relative partial oxygen pressure $\rho$, film thickness $h$ and gap width $\lambda$ are shown in **Fig. 3**. They are calculated for a compressed strain $u_m = -0.25\%$. The values of $T$, $\rho$, $h$ and $\lambda$ are listed for each plot. Similarly to **Fig. 2**, the diagrams calculated for the case of compensated built-in field, $E_{SI} + E_a = 0$, contain the PE phase and three FE phases, FEc, FEaa and FEr. When the $E_{SI} \neq 0$ the PE phase transforms the electret-like FEI phase in thin films. The boundary between the FEI (shown by a sand color) and PE (shown by a russet color) phases is diffuse, especially for a narrow gap ($\lambda = 0.4$ nm). Different parts of **Fig. 3** show how the boundaries of the PE, FEI, and different FE phases depend on the film thickness and gap width.

**Figure 3a** shows the film phase diagram in coordinates $h$ and $\rho$, calculated for the room temperature and narrow gap $\lambda = 0.4$ nm. Here the PE phase is located at $\rho$ very close to unity and film thickness less than the critical, $h < h_{cr}$. The FEc phase is stable for $h > h_{cr}$, and the critical value $h_{cr}$ strongly depends on $\rho$ for thin films, but becomes $\rho$-independent when $h$ exceeds 35 nm (see the vertical dashed line). The FEr phase is stable at $\rho \ll 1$ and film thickness more than the characteristic value $h_{FEr}$, which dependence on $\rho$ is rather weak. The region of FEr phase very slightly increases with $h$ increase.

**Figure 3(b)** is the film phase diagram in coordinates $\lambda$ and $\rho$, calculated for the room temperature and film thickness $h = 25$ nm. The FEc phase is stable for $\lambda < \lambda_{cr}$, and the critical value $\lambda_{cr}$ is $\rho$-independent for narrow gaps, but becomes $\rho$-dependent when $\lambda$ exceeds 0.6 nm (see the vertical dashed line). The FEr phase is stable at $\rho \ll 1$ and gap thickness more than the specific value $\lambda_{FEr}$, which dependence on $\rho$ is rather weak. The region of FEr phase very moderately increases with $\lambda$ increase. Notably, that a relatively simple analytical expression for $\lambda_{cr}$ can be derived from the condition $b_3 = 0$ [see Eq.(8a)]:

$$\lambda_{cr}(T,\rho,h) \cong \frac{2\varepsilon_0\varepsilon_d h\left(\frac{2Q_{12}u_m}{s_{11}+s_{12}}-a_1(T)\right)}{1+h\left(a_1(T)-\frac{2Q_{12}u_m}{s_{11}+s_{12}}\right)\sum_{i=1,2}\frac{\left(eZ_i f_i(T,\rho)\right)^2}{A_i k_B T}}. \quad (13)$$

**Figure 3(c)** is the film phase diagram in coordinates $T$ and $h$, calculated for $\rho = 10^{-6}$ and gap width $\lambda = 0.4$ nm. The FEaa phase is stable at temperatures $T < 300$ K and $h < 22$ nm. The FEaa-FEI boundary is a straight horizontal line, while FEaa-FEr boundary is strongly curved. The FEr phase exists at the temperatures slightly higher than 300 K and film thickness higher than (5 -

20) nm. The FEc phase occupies the widest temperature region (about 200 K in width) between the FEr and FE phase. The FEI phase, which becomes almost indistinguishable from the PE phase with $h$ increase, occupies the high-temperature region. The four phases (FEI, FEc, FEr and FEaa) coexist in a single point.

**Figure 3(d)** is the film phase diagram in coordinates $T$ and $\lambda$ calculated for $\rho = 10^{-6}$ and film thickness $h = 25$ nm. The small region of the FEaa phase, which is stable at temperatures $T < 240$ K and $\lambda > 0.9$ nm, borders with the FEr phase, which occupies the widest region at the diagram. The FEaa-FEr boundary is a steep curve that saturates with $\lambda$ increase. The FEr phase, which is stable in the temperature range between 300 K and 420 K, relatively strongly enlarges its area with $\lambda$ decrease. The FEr-FEc boundary is a smooth curve with a flexion. The FEc phase, which is stable in the temperature range between 300 K and 500 K, has a straight boundary with the FEI phase that is a horizontal line $T = 500$ K. The FEI phase occupies the high-temperature region above 500 K.

Expressions (8a), (9) and (12) explain the $\lambda$- and $h$- size effects of the phase diagram shown in **Fig 3**. In particular, the interplay between the first term $\frac{\lambda h}{\varepsilon_0(\varepsilon_d h + \lambda \varepsilon_{33}^b)}$ in the function $S(T, \rho, h)$ and the last term $\frac{\lambda}{\varepsilon_0(\varepsilon_d h + \lambda \varepsilon_{33}^b)}$ in Eq.(8a) is responsible for the monotonical increase followed by the saturation of $\lambda$- and $h$- dependences of the boundaries between the FEr and FEc phases. Also the expressions describe the behavior of the FEI-FEc boundary at either $h < h_{cr}$ or $\lambda > \lambda_{cr}$ in **Fig. 3(a)** and **Fig. 3(b)**, respectively. These size factors can also explain the shape of the FEI-FEc and FEc-FEr boundaries in **Fig. 3(a)** and **Fig. 3(b)**, respectively. However, the boundaries between the FEaa and FEr phases, which have a complex shape, cannot be described by the abovementioned size dependences, but rather by a complex dependence of the function $S(T, \rho, h)$ on the temperature and pressure.

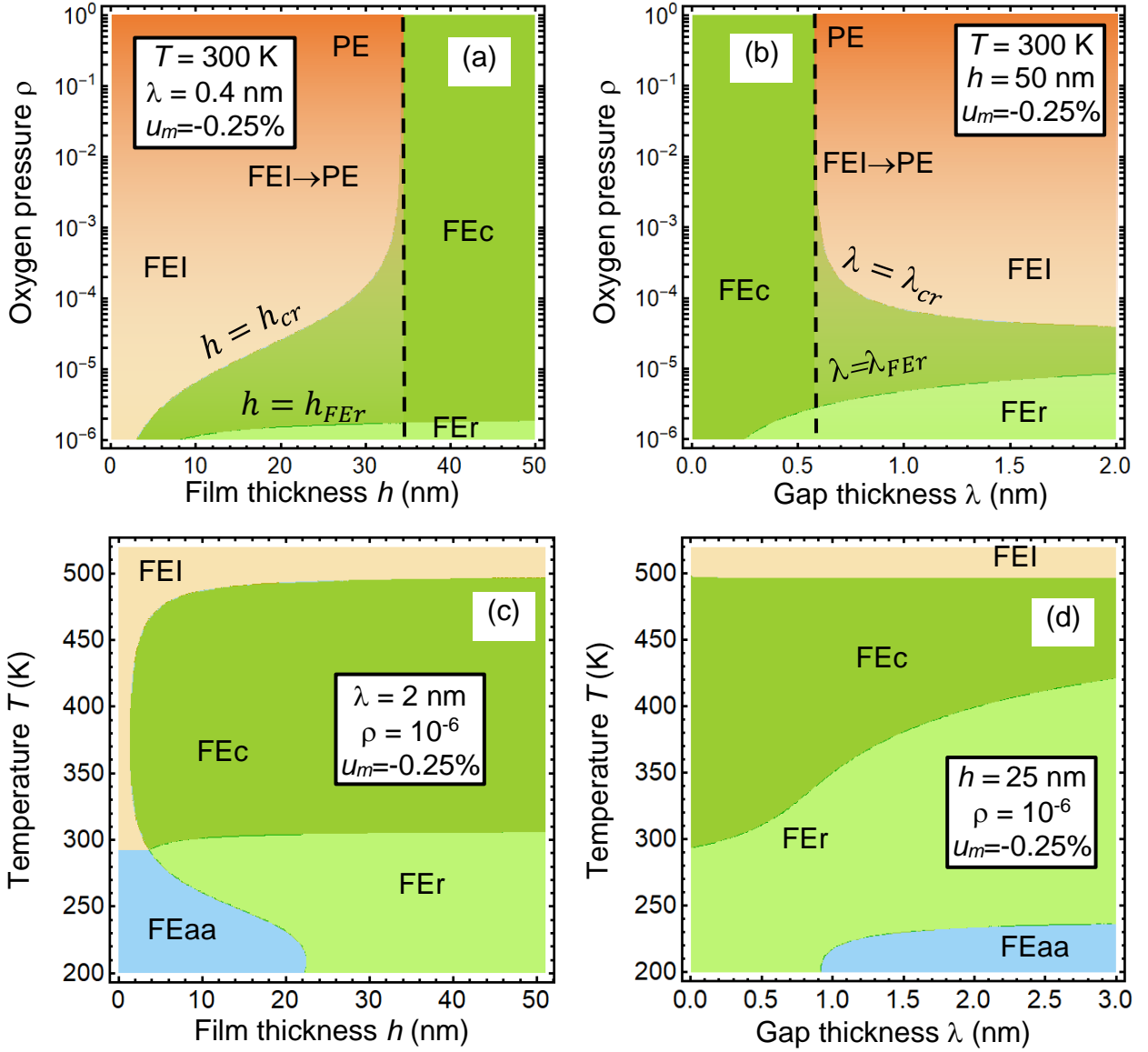

**FIGURE 3.** Typical phase diagrams of a multiaxial FE film in dependence on the temperature $T$ **(a,b)**, relative partial oxygen pressure $\rho$ **(c,d)**, film thickness $h$ **(a,c)** and gap width $\lambda$ **(b,d)**. The values of T, $\rho$, $h$ and $\lambda$ are listed for each plot. The mismatch strain $u_m = -0.25\%$, other parameters are listed in **Tables II-III**.

The diagrams in **Fig. 3** show the possibility to select actual ranges of the film thickness and the gap width for the observation of different FE and FEI phases, and their unusual sequences, including reentrance, and for the control of the phase diagrams, which is most important for various applications. For the material parameters listed in **Tables II-III**, the actual range is $h = (10 - 50)$ nm and $\lambda = (0.4 - 2)$ nm. In experiment the film thickness $h$ and the gap width $\lambda$, and mismatch strain $u_m$ are fixed, while the temperature $T$ and relative partial oxygen pressure $\rho$ are variables. Thus, below we analyze the diagrams in dependence on $T$ and $\rho$ for fixed $h$, $\lambda$ and $u_m$ values within the actual range.

Typical phase diagrams of a multiaxial FE film in dependence on the temperature $T$ and pressure $\rho$ are shown in **Fig. 4** for several film thickness $h$, narrow ($\lambda = 0.4$ nm, the top row) and wider ($\lambda = 2$ nm, the bottom row) gaps. The diagrams in **Fig. S4-S6**, which are calculated for several mismatch strains $u_m$, $\lambda$ and $h$, illustrate the trends shown in **Fig. 4** in more details. For a narrow gap wide regions of the FEr and FEc phases are stable in thick BaTiO$_3$ films. The region of FEI and PE phases are small and located at $\rho$ close to unity [**Fig. 4(a)**]. Under the thickness decrease from 50 nm to 10 nm the FEr phase gradually transforms into the FEaa phase, and the region of FEc phase becomes smaller being substituted by a wider region of FEI phase [compare **Fig. 4(b)** and **4(c)** with **4(a)**].

For a wide gap the large regions of the FEaa and FEI/PE phases are stable in both thick and thin BaTiO$_3$ films [see **Fig. 4(d) – 4(f)**]. Under the thickness decrease from 50 nm to 10 nm the area of FEc and FEr phases decrease, in particular the part of the FEc phase gradually transforms into the FEI phase, and the part of the FEr phase becomes the FEaa phase [compare **Fig. 4(d), 4(e)** and **4(f)**].

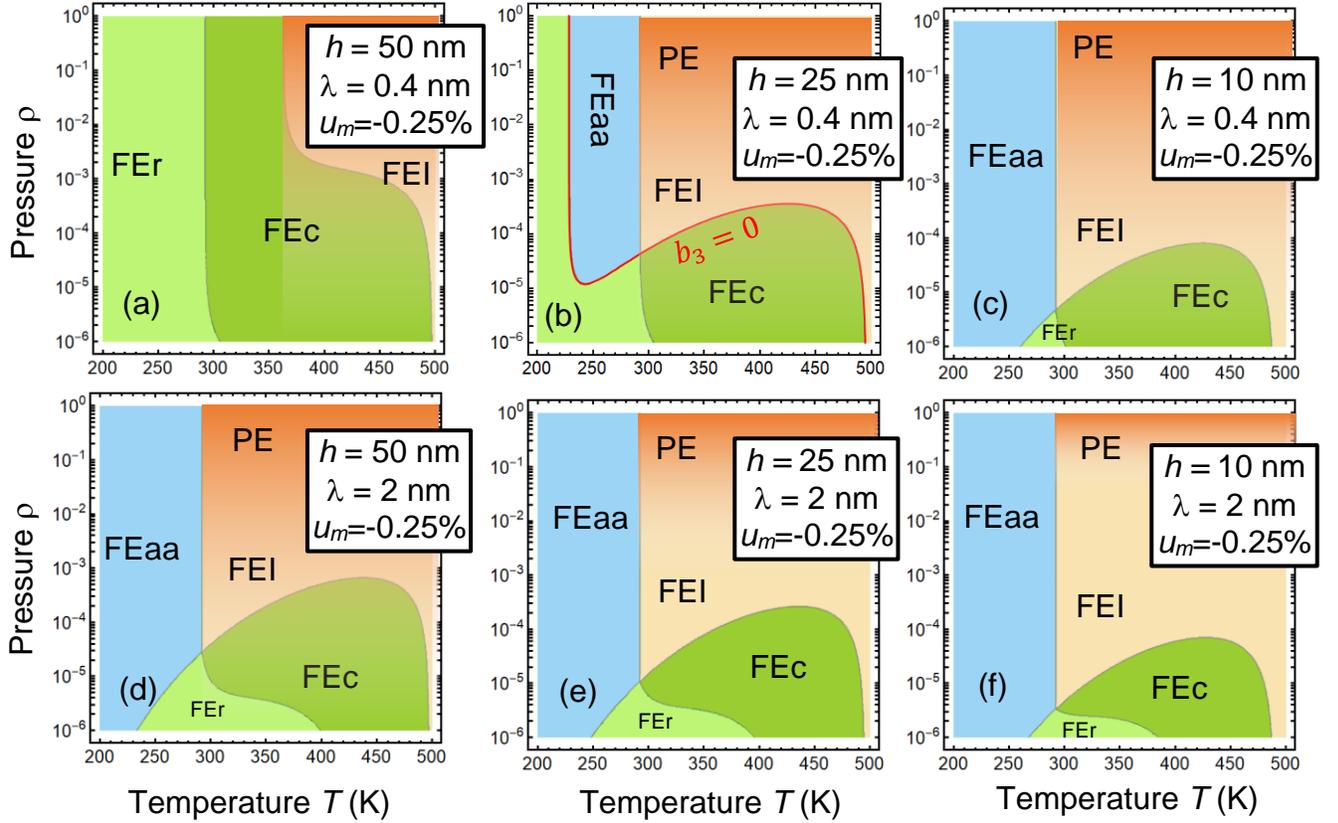

**FIGURE 4.** Typical phase diagrams of a multiaxial FE film in dependence on the temperature $T$ and relative partial oxygen pressure $\rho$. The values of $h$ and $\lambda$ are listed for each plot. A red curve in the plot (b) corresponds to the condition $b_3 = 0$. The mismatch strain $u_m = -0.25\%$, other parameters are listed in **Tables II-III.**

The common features of the diagrams shown in **Fig. 4(b)-4(f)** are the unusual parabolic-like shape of the reentrant FEc phase region, the straight boundary between the FEI/PE and FEaa phases, the curved boundary between FEc and FEr phases, and a single point, where all stable phases (FEI, FEc, FEr and FEaa) coexist. All these features originate from the surface screening by oxygen ions. Mathematically, the features are defined by the dependences of the coefficients $b_3$, $b_{33}$ and $b_{13}$ on the variables $T$ and $\rho$ (see Eqs.(8) for details). As it was discussed above, the coefficients contain the function $S(T, \rho, h)$, which diverges at $T \to 0$ and monotonically decreases with $T$ increase. In result the coefficient $b_3$ nonmonotonically depends on $T$ and can change its several times at fixed $\rho$ value. The condition $b_3 = 0$ gives us the "reentrant" boundaries between FEI and FEc phases, FEr and FEaa phases; an example is shown by a red line in **Fig. 4(b)**.

## D. Quasi-static field dependences of polarization, dielectric permittivity and piezoelectric coefficients

To illustrate the polar, dielectric and piezoelectric properties of the FEI, FEc, FEr and FEaa phases, which can be stable in thin multiaxial FE films covered with ions, we calculated and analyzed corresponding quasi-static dependences on the applied voltage $U$ using Eqs.(11). When calculating the dependences of polarization $P(U)$ on the periodic applied voltage, $U = U_0 \sin(\omega t)$, we introduce the dimensionless frequency $w = \omega \tau$, which govern the polarization response to an external field, and the dimensionless characteristic time $\tau = \frac{\Gamma_P}{2|\alpha_p|}$. The criterium of low frequency is $w \ll 1$. Typical results are shown in **Figs. 5** and **6** for the frequency $w = 10^{-3}$. Since we regard that $\varepsilon_d h \gg \lambda \varepsilon_{33}^b$, the acting field $E_a(U, h) = -\frac{\varepsilon_d U}{\varepsilon_d h + \lambda \varepsilon_{33}^b}$ is approximately equal to the external electric field $E_e = -\frac{U}{h}$ (see Eq.(10c) for details).

Quasi-static dependences of the out-of-plane polarization $P_3$, dielectric permittivity $\varepsilon_{33}$ and piezoelectric coefficient $d_{33}$ on the external electric field $E_e$ are shown in **Fig. 5** for several values of the film thickness and partial oxygen pressures. The parameters $\lambda = 2$ nm, $u_m = -0.25\%$ and $T = 350$ K are chosen in such way, that the hysteresis behavior of polarization, double maxima of permittivity and piezocoefficient exist for the pressures far from unity, $\rho = (10^{-6} - 10^{-5})$, and disappear for the pressures closer to the unity, $\rho = (10^{-4} - 1)$ (compare different curves in **Fig. 5**). The hysteresis-less dependences are left-shifted due to the built-in field $E_{SI}$, and the shift significantly increases with the film thickness decrease, as explained by the size factor $\frac{\lambda}{\varepsilon_0(\varepsilon_d h + \lambda \varepsilon_{33}^b)}$ in Eq.(10b) for the $E_{SI}$. The shift is the most pronounced indicator of the FEI phase. However, the shift exists and becomes noticeable for the FE hysteresis loops calculated at $\rho \ll 10^{-4}$ when the film becomes thinner (compare black and red loops in **Fig. 5**).

It is important for applications, that $d_{33}$ is significantly enhanced for $\rho = (10^{-6} - 10^{-5})$ in the region of coercive fields [see red and black curves in Fig.**5(g)-(i)**]. To quantify the enhancement effect, we need to consider (or exclude) the role of the domain formation in the presence of surface ions. Indeed, the domain formation can appear in the multiaxial FE films [64, 65, 66, 67] under incomplete screening of their polarization. In the considered case the domain formation is likely for the near-equilibrium oxygen pressures $(10^{-2} < \rho < 10^2)$ and wide gaps ($\lambda > 1$ nm). However, the case $\rho \to 1$ is not of our interest, because the reentrant phases and enhance polar properties correspond to $\rho = (10^{-6} - 10^{-5})$.

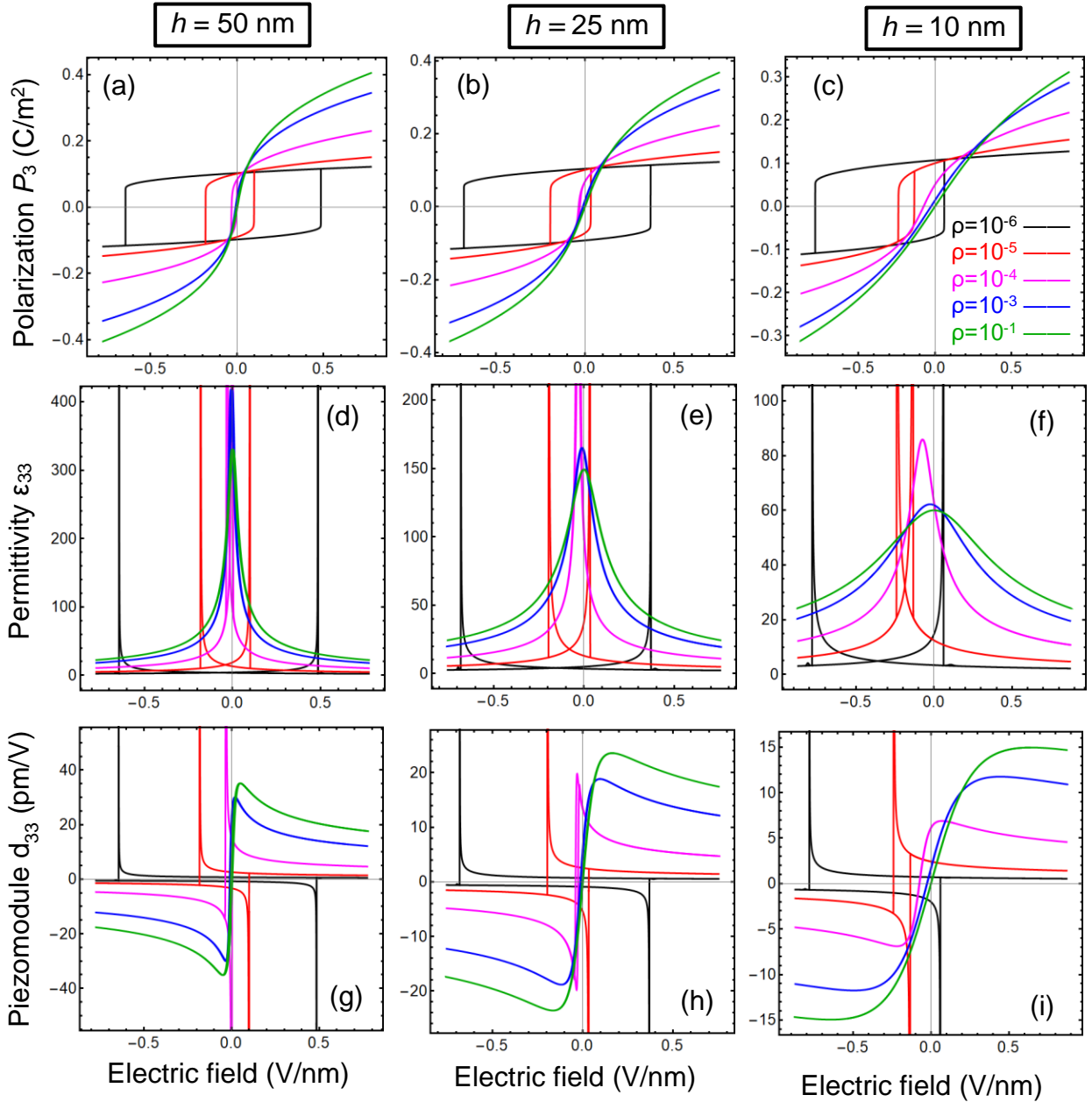

**FIGURE 5.** Quasi-static dependences of the out-of-plane polarization $P_3$ (**a, b, c**), dielectric permittivity $\varepsilon_{33}$ (**d, e, f**) and piezoelectric coefficient $d_{33}$(**g, h, i**) on external electric field $E_e$ calculated for several values of the film thickness 50 nm (**a, d, g**), 25 nm (**b, e, h**) and 10 nm (**c, f, i**) and partial oxygen pressures $\rho = 10^{-6}$ (black curves), $10^{-5}$ (red curves), $10^{-4}$ (magenta curves), $10^{-3}$ (blue curves) and $10^{-1}$ (green curves). Parameters: $\lambda = 2$ nm, $u_m = -0.25\%$ and $T = 350$ K; other parameters are listed in **Tables II-III**.

The field dependences of the in-plane polarization components $P_1$ and $P_2$, dielectric permittivity $\varepsilon_{13}$ and $\varepsilon_{23}$, and piezoelectric coefficient $d_{13}$ and $d_{23}$, are shown in **Fig. S8** for the film thickness $h = 50, 25$ and 10 nm and relative pressure $\rho = 10^{-6}$. The dependences are calculated for the same parameters as the dependences in **Fig. 5**. The purpose of **Figs. S8** is to illustrate the butterfly-like shape of $P_1$ and $P_2$ hysteresis loops in the FEaa and FEr phases, specific hysteresis of $d_{13}$ and $d_{23}$ in the phases, and their changes with the thickness decrease and oxygen pressure variation. The butterfly-like $P_{1,2}$-loops, as well as $d_{13}$ and $d_{23}$ loops, are also shifted by the built-in field $E_{SI}$, and the shift increases with the film thickness decrease.

In complex, results presented in **Fig. 5** and **S8**, point on the possibility to control the appearance and features of ferroelectric, dielectric and piezoelectric hysteresis in multiaxial FE films covered by surface ions by varying their concentration via the partial oxygen pressure.

Quasi-static dependences of the out-of-plane polarization $P_3$, dielectric constant $\varepsilon_{33}$ and piezoelectric coefficient $d_{33}$ on the external electric field $E_e$ are shown in **Fig. 6** for several values of the film thickness and temperatures, $\lambda = 2$ nm, $u_m = -0.25\%$ and $\rho = 10^{-6}$. The parameters are chosen in such way, that the hysteresis behavior of polarization, double maxima of permittivity and piezocoefficient exist in the temperature range (290 – 480) K for thin films and in a slightly wider range (250 – 500) K for thicker films. The temperature range mainly corresponds to the reentrant FEc phase, and so the coercive field is maximal in the middle of the range (compare different curves in **Fig. 6**). Both, hysteresis-less curves and hysteresis loops are left-shifted due to the built-in field $E_{SI}$, and the shift significantly increases with the film thickness decrease and temperature increase. The shift of the hysteresis-less curves is the most pronounced indicator of the FEI phase (compare black curves for different film thickness).

The field dependences of in-plane polarization components $P_1$ and $P_2$, dielectric permittivity $\varepsilon_{13}$ and $\varepsilon_{23}$, and piezoelectric coefficient $d_{13}$ and $d_{23}$, are shown in **Figs. S9-11** for the film thickness $h = 50, 25$ and 10 nm. The dependences are calculated for the same parameters as the dependences in **Fig. 6**. The purpose of **Figs. S9-11** is to illustrate the transformations of $P_1$ and $P_2$ E-dependences, which happen in the FEI, FEaa and FEr phases with changes of the film thickness and temperature.

Generally, the results presented in **Fig. 5** and **S9-11**, point on the temperature control of the appearance and features of ferroelectric, dielectric and piezoelectric hysteresis in multiaxial FE films covered by surface ions.

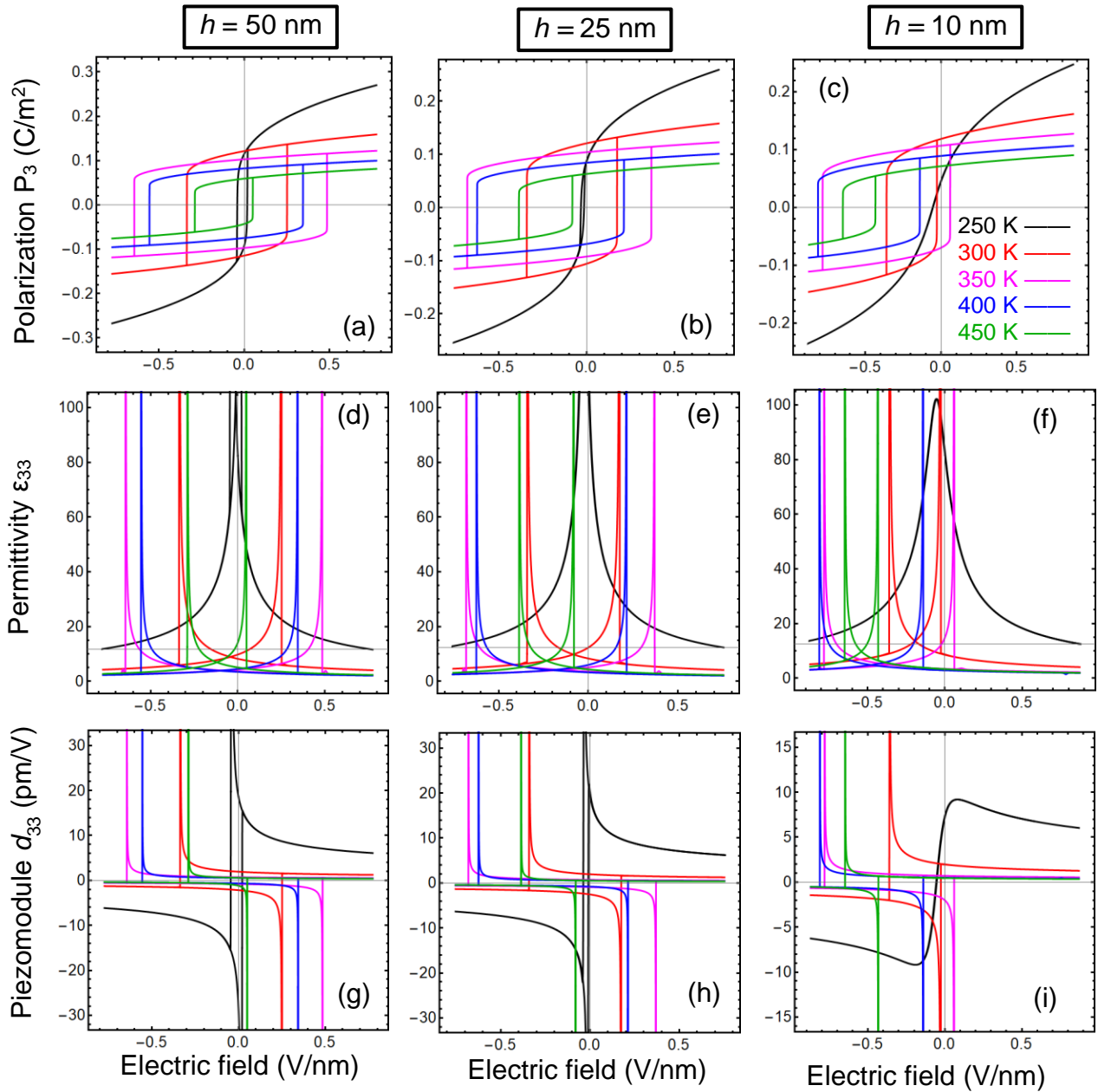

**FIGURE 6.** Quasi-static dependences of the out-of-plane polarization $P_3$ **(a, b, c)**, dielectric constant $\varepsilon_{33}$ **(d, e, f)** and piezoelectric coefficient $d_{33}$ **(g, h, i)** on external electric field $E_e$ calculated for several values of film thickness 50 nm **(a, d, g)**, 25 nm **(b, e, h)** and 10 nm **(c, f, i)** and temperatures $T = 250$ (black curves),

300 (red curves), 350 (magenta curves), 400 (blue curves) and 450 K(green curves). Parameters $\lambda$ =2 nm, $u_m = -0.25\%$ and $\rho =10^{-6}$; other parameters are listed in **Tables II-III.**

### E. Multi-objective Bayesian exploration with Gaussian Process

Complementary we provide detail investigation of both phase and minimum free energy of a multiaxial FE film in the dependence of the temperature $T$ and mismatch strain $u_m$, as shown in **Fig. 7** with further detailed figures and formulations as explained in **Appendix D.** The collab code is available in [68].

We started with configuring the ground states of both functionalities, using low-sampling (10x10) based exhaustive grid search. Since this is computationally costly, and to further analyze the domain space, we develop a multi-objective Bayesian optimization (**MOBO**) architecture where it first replaces with cheap surrogate models (e.g., Gaussian process - GP) for each chosen functionality and then attempt for rapid exploration/exploitation towards learning interesting domain region through sequential sampling from optimizing a multi-function-based acquisition function. To date, there are only few applications of MOBO for material discovery in optimizing multiple target properties, considering both experimental and computer simulated data [69]. MOBO has also been applied for efficient discovery of the targeted materials, performed by a thermodynamically-consistent micromechanical model that predicts the materials response based on its composition and microstructure [70]. Recently, a physics driven MOBO is implemented for the discovery of optimal Pareto frontier between the energy storage and loss in interfacial-controlled ferroelectric materials [49]. Unlike these applications, here the goal is to learn the overall parameter space and thus, we are not optimizing the functions but want to learn the overall trend of the functionalities. Thus, we build our acquisition function, suitable for full exploration of both the functionalities (phase and free energy) jointly. For all the analysis in **Fig. 7**, the MOBO is started with evaluating 20 randomly selected locations (samples) from the 2-4 LGD model using material parameters of a bulk BaTiO$_3$ and stopped after 230 MOBO guided evaluations from a dense 50x50 map, with a total of 250 expensive evaluations.

**Figs. 7(a)** and **7(b)** are the ground state phase diagram and free energy calculated using low-sampling (10x10) exhaustive grid search for the same parameter space as in **Fig. 1(b)**. As anticipated the diagram is a low-resolution analog of **Fig. 1(b)**. We observe as four distinct phases: PE, FEc, FEaa and FEr. **Figs. 7(c)** and **7(d)** are GP predicted dense (50x50) maps of phase and free energy, respectively. Here the function evaluations were done at the locations represented by filled dots in the figures. Thus, we build the dense map with only 10% evaluation of total grids (250 out 2500). With $u_m(\%)$ close towards zero and with high (low) values of $T$, we find the PE (FEr) phase(s). As the absolute value of $u_m$ increases the phase changes to FEaa or FEc, respectively. For the respective free energy map, we obtain a higher energy value for the PE and FEr phases, lower energy towards FEaa and FEc phases. **Figs. D1(a, b)** in supplementary are the uncertainty maps for respective GP predicted maps in **Figs. 7(c, d)**.

**Figs. 7(e)** and **7(f)** are the ground state phase diagram and free energy calculated using low-sampling exhaustive grid search for the same parameter space as in **Fig. 2(a)**; and the diagram is a low-resolution analog of **Fig. 2(a)**. **Figs. 7(g)** and **7(h)** are the GP predicted dense map of phase and free energy, respectively. With $u_m(\%)$ close to zero and with high (or low) values of $T$, we find the PE (or FEr) phase. As the value of $u_m$ increases in the positive (or negative) direction, we see phase changes to FEaa (or FEc). The trend is similar to the phase map in **Fig. 7(d)**, however the trend in the energy map is very different to **Fig. 7(g)**. Here, we obtain higher energy value for the PE, FEaa and FEr phases and lower energy towards FEc phase. **Figs. D1(c, d)** in supplementary are the uncertainty maps for respective GP predicted maps in **Figs. 7(g, h)**.

**Figs. 7(i)** and **7(j)** are the ground state phase diagram and free energy calculated using low-sampling exhaustive grid search for the same parameter space as in **Fig. 2(f)**; and the diagram is a low-resolution analog of **Fig. 2(f)**. **Figs. 7(k)** and **(l)** are the GP predicted dense map of phase and free energy, respectively. With $u_m(\%)$ close towards zero and with high values of $T$, we find the PE phase We have another region of the PE phase where both values, $T$ and $u_m$, are lower. We find the FEaa phase mostly as $u_m$ increases in the positive direction and sometimes when $u_m$ is close to zero. Just like the PE phase, the reentrant FEc phase also has two distinct regions mostly as $u_m$ increases in the negative direction, where these two FEc regions are separated by the region of the FEr phase. Here, we obtain higher energy value for the PE, FEaa phases and lower energy towards the FEc and FEr phases. **Figs. D1(e, f)** in Suppl.Mat.[54] are the uncertainty maps for respective GP predicted maps in **Figs. 7(k, l)**.

To resume, the LGD-SH description of a multiaxial FE film appears very promising for MOBO, being important for the treatment of potential experiments.

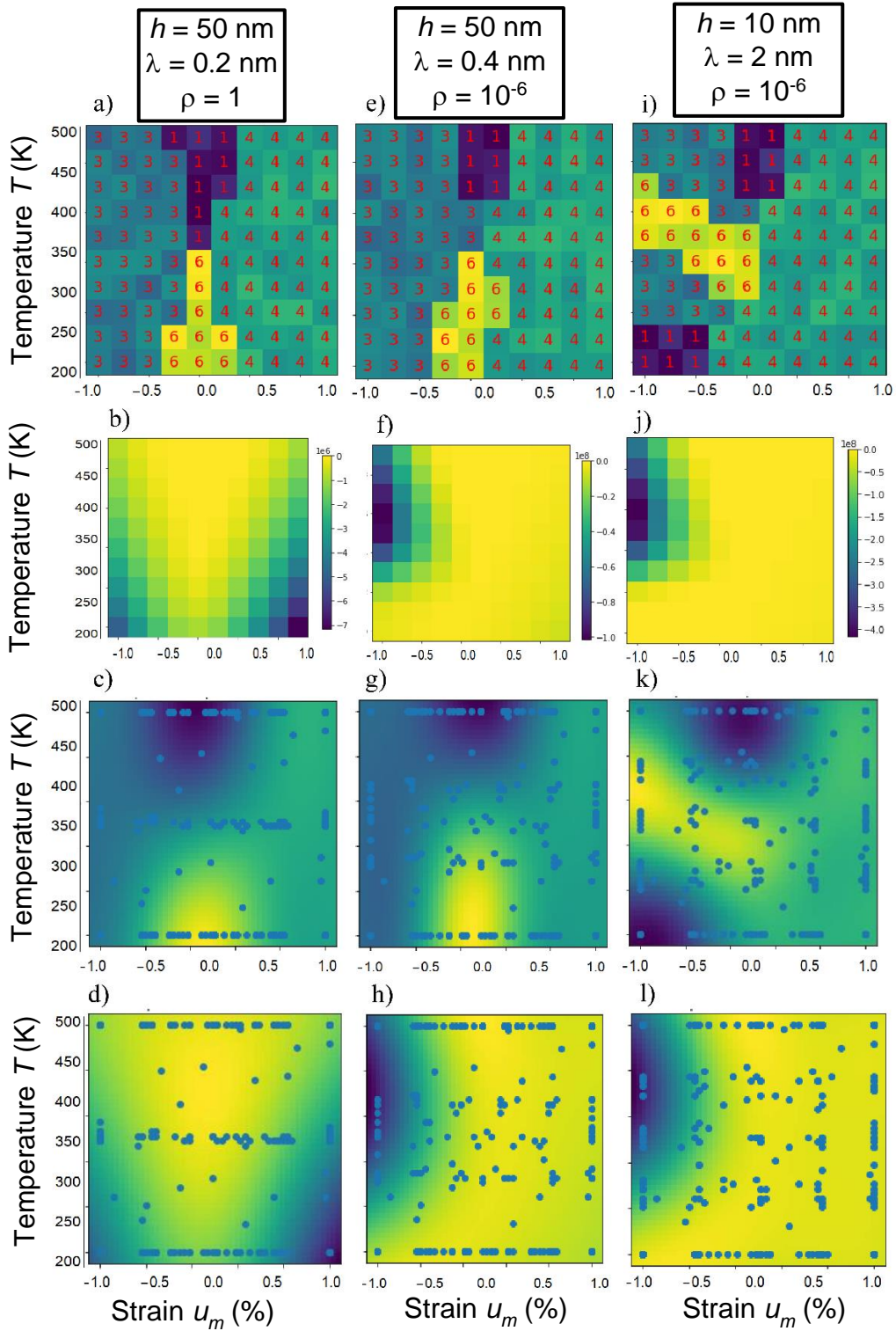

**FIGURE 7**. Joint exploration of the phase and minimum free energy diagrams of a multiaxial FE film in dependence of the temperature $T$ and mismatch strain $u_m$. Parameter space: $T(K) = [200, 500]$ and $u_m(\%) = [-1, 1]$. The values of $h, \lambda, \rho$ are same for each column corresponding to the parameter space for **Figs. 1(b), 2(a)** and **2(f)**, respectively. Maps **(a, e, i)** are the ground states of phase diagrams and maps **(b, f, j)** are the ground states of minimum free energy diagrams calculated using low-sampling exhaustive grid search; maps **(c, g, k)** are the predicted phases dense maps of ground truth maps **(a, e, i)**, and the maps **(d, h, l)** are the predicted energy dense maps of ground truth maps **(b, f, j)**. The filled dots in the predicted maps are the locations only where the function evaluation is done to build the respective dense maps of the entire domain space. The numbers printed within the grids of the ground state phase diagrams are: PE – 1, FEa – 2, FEc – 3, FEaa – 4, FEac – 5, FEr – 6. Actually, FEa and FEac phases are absent. Other parameters are listed in **Table II-III**.

## V. CONCLUSION

Using the combined LGD-SH approach we described the electrochemical switching and rotation of polarization vector in a ferroelectric film covered by surface ions with a charge density proportional to the relative partial oxygen pressure. We calculate the phase diagrams and analyze the dependence of polarization components on the applied voltage, and discuss the peculiarities of quasi-static ferroelectric, dielectric and piezoelectric hysteresis loops in thin strained multiaxial ferroelectric films.

The nonlinear surface screening by oxygen ions makes the phase diagrams very different from the known ones, with the most interesting feature being the appearance of the ferroelectric reentrant phases. The common features of the diagrams are the unusual parabolic-like shape of the reentrant FEc phase region, the straight boundary between the FEI/PE and FEaa phases, the curved boundary between FEc and FEr phases, and a single point, where all stable phases (FEI, FEc, FEr and FEaa) coexist. The LGD-SH description of a multiaxial FE film appears very promising for MOBO, being important for the treatment of potential experiments.

Both, hysteresis-less curves and hysteresis loops of polarization vector and piezoelectric tensor components are shifted due to the built-in field induced by the surface ions, and the shift significantly increases with the film thickness decrease and temperature increase. The shift of the hysteresis-less curves is the most pronounced indicator of the FEI phase. Obtained results point on the possibility to control the appearance and features of ferroelectric, dielectric and piezoelectric hysteresis in multiaxial FE films covered by surface ions by varying their concentration via the partial oxygen pressure.

We also predict the enhanced properties at the polarization rotation, and to quantify the effect, we need to explore the role of the domain formation in the presence of surface defects, which can appear in the multiaxial FE films [64-66] under the incomplete screening conditions. In the considered case the domain formation is likely for the near-equilibrium relative oxygen pressures and wide gaps. However, the region of pressures is far from our interest, because the reentrant phases and enhance polar properties correspond to high oxygen excess or deficiency. Not less important can be the interplay between proximity to polarization rotation boundary and emergence of nanoscale phase separation [71].

**Acknowledgements.** This effort is based upon work supported by the U.S. Department of Energy, Office of Science, Office of Basic Energy Sciences Energy Frontier Research Centers program under Award Number DE-SC0021118 (S.V.K. and A.B.) and performed at the Oak Ridge National Laboratory's Center for Nanophase Materials Sciences (CNMS), a U.S. Department of Energy, Office of Science User Facility. A.N.M. and N.V.M. work is supported by the National Academy of Sciences of Ukraine. A.N.M and H.V.S. work is supported by the National Research Foundation of Ukraine (Grant applications Φ81/41481 and 2020.02/0027).

**Authors' contribution.** S.V.K. and A.N.M. generated the research idea. A.N.M. proposed the theoretical model, derived analytical results and interpreted numerical results, obtained by E.A.E and H.V.S. A.B. performed MOBO analysis. A.N.M. and S.V.K. wrote the manuscript draft. All co-authors worked on the results discussion and manuscript improvement.

# SUPPLEMENTARY MATERIAL
## APPENDIX A. Dependence of surface charge on oxygen ions

Using Eq.(22) from the SH paper [43]:

$$\frac{\theta}{1-\theta} = \rho^{-1/n_i} \exp\left(-\frac{\Delta G_i^{00}+eZ_i\phi}{k_B T}\right), \quad (A.1)$$

we obtain that

$$\theta_i(\phi) = \left(1 + \rho^{1/n_i} \exp\left(\frac{\Delta G_i^{00}+eZ_i\phi}{k_B T}\right)\right)^{-1}. \quad (A.2)$$

Next using Eq.(24) from the SH paper [43], we obtain Eq.(4) from the main text:

$$\sigma_0[\phi] = \sum_{i=1}^{2} \frac{eZ_i \theta_i(\phi)}{A_i} \equiv \sum_{i=1}^{2} \frac{eZ_i}{A_i}\left(1 + \rho^{1/n_i} \exp\left(\frac{\Delta G_i^{00}+eZ_i\phi}{k_B T}\right)\right)^{-1}. \quad (A.3)$$

## APPENDIX B. Free energy with renormalized coefficients

Below we assume that the polarization distribution $P_i(x, y, z)$ is smooth enough, the coupled nonlinear algebraic equations for the polarization averaged over film thickness and surface charge density $\sigma$ are valid:

$$\Gamma\frac{\partial P_3}{\partial t} + \left(\tilde{a}_3 + \tilde{a}_{33}P_3^2 + \tilde{a}_{13}(P_1^2 + P_2^2)\right)P_3 = \frac{\Psi(U,\sigma,P_3)}{h}, \quad (B.1a)$$

$$\Gamma\frac{\partial P_1}{\partial t} + (\tilde{a}_1 + \tilde{a}_{11}P_1^2 + \tilde{a}_{12}P_2^2 + \tilde{a}_{13}P_3^2)P_1 = 0, \quad (B.1b)$$

$$\Gamma\frac{\partial P_2}{\partial t} + (\tilde{a}_1 + \tilde{a}_{11}P_2^2 + \tilde{a}_{12}P_1^2 + \tilde{a}_{13}P_3^2)P_2 = 0, \quad (B.1c)$$

$$\tau\frac{\partial \sigma}{\partial t} + \sigma = \sigma_0[\Psi(U,\sigma,P_3)]. \quad (B.1d)$$

Here $\tilde{a}_3(T,\rho,u_m) = a_1(T) - u_m\frac{4Q_{12}}{s_{11}+s_{12}}$, $\tilde{a}_1(T,u_m) = a_1(T) - 2u_m\frac{Q_{11}+Q_{12}}{s_{11}+s_{12}}$, $\tilde{a}_{11} = a_{11} + 4\frac{s_{11}(Q_{11}^2+Q_{12}^2)-2Q_{11}Q_{12}s_{12}}{2(s_{11}^2-s_{12}^2)}$, $\tilde{a}_{33} = a_{11} + \frac{4Q_{12}^2}{s_{11}+s_{12}}$, $\tilde{a}_{12} = a_{12} - 4\frac{s_{12}(Q_{11}^2+Q_{12}^2)-2Q_{11}Q_{12}s_{11}}{s_{11}^2-s_{12}^2} + 2\frac{Q_{44}^2}{s_{44}}$, and $\tilde{a}_{13}(T,\rho,h) = a_{12} + 4\frac{Q_{12}(Q_{11}+Q_{12})}{s_{11}+s_{12}}$ are already renormalized by mismatch strain.

The overpotential in Eq.(B.1d) is given by expression $\Psi(U,\sigma,P_3) = \frac{\lambda(\sigma-P_3)+\varepsilon_0\varepsilon_d U}{\varepsilon_0(\varepsilon_d h+\lambda\varepsilon_{33}^b)}h$ and the function $\sigma_0[\psi] = \sum_i \frac{eZ_i\theta_i(\psi)}{A_i} \equiv \sum_i \frac{eZ_i}{A_i}\left(1 + \rho^{1/n_i} \exp\left(\frac{\Delta G_i^{00}+eZ_i\psi}{k_B T}\right)\right)^{-1}$. The electric potential acting in the dielectric gap ($\phi_d$) and in the ferroelectric film ($\phi_f$) linearly depends on the coordinate z and overpotential, namely $\phi_d = U - \frac{z+\lambda}{\lambda}(U - \Psi)$ and $\phi_f = (h-z)\frac{\Psi}{h}$. Below we'll use the designations

$$\Delta G_i^{0p} = \Delta G_i^{00} + \frac{k_B T}{n_i}\ln(\rho). \quad (B.1e)$$

Next we consider the stationary case, when one can put $\sigma = \sigma_0[\Psi(U,\sigma,P_3)]$ in Eqs.(B.1d). Corresponding free energy $G[P_3,\Psi]$ that's formal minimization, $\frac{\partial G[P_3,\Psi]}{\partial P_3} = 0$ and $\frac{\partial G[P_3,\Psi]}{\partial \Psi} = 0$, leads to Eqs.(B.1), has the form:

$$\frac{G[P_3,\Psi]}{S} = h\left(\frac{\tilde{a}_3}{2}P_3^2 + \frac{\tilde{a}_1}{2}(P_1^2 + P_2^2) + \frac{\tilde{a}_{11}}{4}(P_1^4 + P_2^4) + \frac{\tilde{a}_{33}}{4}P_3^4 + \frac{\tilde{a}_{12}}{2}P_1^2 P_2^2 + \frac{\tilde{a}_{13}}{2}(P_1^2 + P_2^2)P_3^2\right) -$$
$$\Psi P_3 - \varepsilon_0 \varepsilon_{33}^b \frac{\Psi^2}{2h} - \frac{\varepsilon_0 \varepsilon_d}{2}\frac{(\Psi-U)^2}{\lambda} + \int_0^\Psi \sigma_0[\varphi]d\varphi, \qquad (B.2a)$$

The energy (B.2a) has absolute minima at high $\Psi$ values. So, according to the Biot's variational principle, let us find for the incomplete thermodynamic potential, which partial minimization over $P_i$ will give the equations of state, and, at the same time, it has an absolute minimum at finite $P_i$ values. Substituting here the expression for the overpotential $\frac{\Psi}{h} = \frac{\lambda(\sigma_0[\Psi] - P_3) + \varepsilon_0 \varepsilon_d U}{\varepsilon_0(\varepsilon_d h + \lambda \varepsilon_{33}^b)}$ we derived the single equation for the average polarization:

$$(\tilde{a}_3 + \tilde{a}_{33}P_3^2 + \tilde{a}_{13}(P_1^2 + P_2^2))P_3 = \frac{\lambda}{\varepsilon_0(\varepsilon_d h + \lambda \varepsilon_{33}^b)}\sum_{i=1,2}\frac{eZ_i}{A_i}\left(1 + exp\left[\frac{\Delta G_i^{op} + eZ_i h(\tilde{a}_3 + \tilde{a}_{33}P_3^2 + \tilde{a}_{13}(P_1^2 + P_2^2))P_3}{k_B T}\right]\right)^{-1} - \frac{\lambda P_3 - \varepsilon_0 \varepsilon_d U}{\varepsilon_0(\varepsilon_d h + \lambda \varepsilon_{33}^b)} \qquad (B.3)$$

Corresponding potential, which minimization over $P_3$ gives Eq.(B.3), has the form:

$$F[P_i] = \left(\frac{\lambda}{2\varepsilon_0(\varepsilon_d h + \lambda \varepsilon_{33}^b)} + \frac{\tilde{a}_3}{2}\right)P_3^2 + \frac{\tilde{a}_1}{2}(P_1^2 + P_2^2) + \frac{\tilde{a}_{11}}{4}(P_1^4 + P_2^4) + \frac{\tilde{a}_{33}}{4}P_3^4$$
$$+ \frac{\tilde{a}_{12}}{2}P_1^2 P_2^2 + \frac{\tilde{a}_{13}}{2}(P_1^2 + P_2^2)P_3^2 + \frac{\varepsilon_d U P_3}{\varepsilon_d h + \lambda \varepsilon_{33}^b} - \qquad (B.4)$$
$$- \frac{\lambda}{\varepsilon_0(\varepsilon_d h + \lambda \varepsilon_{33}^b)}\sum_{i=1,2}\frac{eZ_i}{A_i}\int_0^{P_3} dp\left(1 + exp\left(\frac{\Delta G_i^{op} + eZ_i h(\tilde{a}_3 + \tilde{a}_{33}p^2 + \tilde{a}_{13}(P_1^2 + P_2^2))p}{k_B T}\right)\right)^{-1}$$

Next, we consider the condition $\left|\frac{eZ_i \Psi}{k_B T}\right| \ll 1$ and use that

$$\left(1 + exp\left(\frac{\Delta G_i^{op} + eZ_i \Psi}{k_B T}\right)\right)^{-1} \approx f_i(T,\rho)\left(1 - \frac{eZ_i \Psi}{k_B T}f_i(T,\rho)\right), \qquad (B.5a)$$

were we introduced the sort of "level filling factor":

$$f_i(T,\rho) = \left(1 + exp\left(\frac{\Delta G_i^{op}}{k_B T}\right)\right)^{-1} \qquad (B.5b)$$

In this case the last term in the expression (B.4) can be further simplified as

$$-\frac{\lambda}{\varepsilon_0(\varepsilon_d h + \lambda \varepsilon_{33}^b)}\sum_{i=1,2}\frac{eZ_i}{A_i}\int_0^{P_3} dp\left(1 + exp\left(\frac{\Delta G_i^{op} + eZ_i h(\tilde{a}_3 + \tilde{a}_{33}p^2 + \tilde{a}_{13}(P_1^2 + P_2^2))p}{k_B T}\right)\right)^{-1} \approx \qquad (B.6)$$
$$-\frac{\lambda}{\varepsilon_0(\varepsilon_d h + \lambda \varepsilon_{33}^b)}\sum_{i=1,2}\frac{eZ_i}{A_i}f_i(T,\rho)\int_0^{P_3} dp\left(1 - f_i(T,\rho)\frac{eZ_i h}{k_B T}(\tilde{a}_3 + \tilde{a}_{33}p^2 + \tilde{a}_{13}(P_1^2 + P_2^2))p\right)$$

Using that

$$\int_0^{P_3} dp \left(1 - f_i(T,\rho)\frac{eZ_ih}{k_BT}\left(\tilde{a}_3 + \tilde{a}_{33}p^2 + \tilde{a}_{13}(P_1^2 + P_2^2)\right)p\right) = P_3 - f_i(T,\rho)\frac{eZ_ih}{k_BT}\left(\frac{\tilde{a}_3}{2}P_3^2 + \frac{\tilde{a}_{33}}{4}P_3^4 + \frac{\tilde{a}_{13}}{2}(P_1^2 + P_2^2)P_3^2\right) \quad (B.7)$$

One could get from Eqs.(B.4) and (B.6) that

$$F[P_i] = \left(\frac{\lambda}{2\varepsilon_0(\varepsilon_d h + \lambda\varepsilon_{33}^b)} + \frac{\tilde{a}_3}{2}(1+S)\right)P_3^2 + \frac{\tilde{a}_1}{2}(P_1^2 + P_2^2) + \frac{\tilde{a}_{11}}{4}(P_1^4 + P_2^4) + \frac{\tilde{a}_{33}}{4}(1+S)P_3^4$$
$$+ \frac{\tilde{a}_{12}}{2}P_1^2P_2^2 + \frac{\tilde{a}_{13}}{2}(1+S)(P_1^2+P_2^2)P_3^2 - \left(\frac{\lambda}{\varepsilon_0(\varepsilon_d h + \lambda\varepsilon_{33}^b)}\sum_{i=1,2}\frac{eZ_i}{A_i}f_i(T,\rho) - \frac{\varepsilon_d U}{\varepsilon_d h + \lambda\varepsilon_{33}^b}\right)P_3 \quad (B.8)$$

The function $S(T,\rho,h)$ in Eqs.(B.8) is

$$S(T,\rho,h) = \frac{\lambda h}{\varepsilon_0(\varepsilon_d h + \lambda\varepsilon_{33}^b)}\sum_{i=1,2}\frac{(eZ_if_i(T,\rho))^2}{k_BTA_i} \quad (B.9a)$$

The effective field, produced by ionic charge and applied potential $U$, has the form:

$$E_{eff}(U,\Delta G_i^{00}) = \frac{\lambda}{\varepsilon_0(\varepsilon_d h + \lambda\varepsilon_{33}^b)}\sum_{i=1,2}\frac{eZ_i}{A_i}f_i(T,\rho) - \frac{\varepsilon_d U}{\varepsilon_d h + \lambda\varepsilon_{33}^b}, \quad (B.9b)$$

# APPENDIX C. Auxiliary figures
## A. Phase diagrams in coordinates temperature – mismatch strain

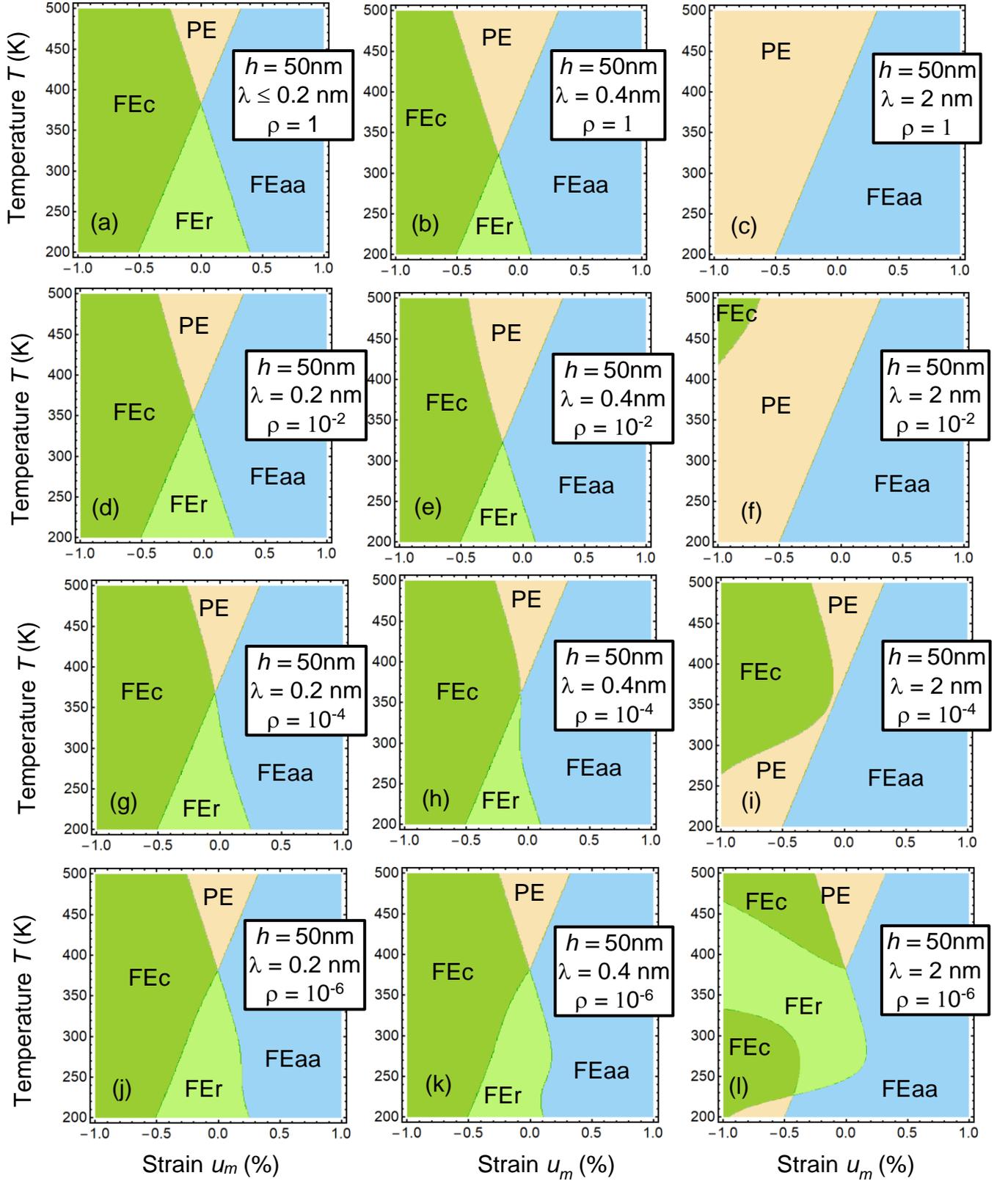

**FIGURE S1.** Typical phase diagrams of a multiaxial FE film in dependence on the temperature $T$ and mismatch strain $u_m$. The values of relative partial oxygen pressure, $\rho = 1, 10^{-2}, 10^{-4}, 10^{-6}$, and gap thickness, $\lambda = 0.2, 0.4, 2$ nm, are listed for each plot. The diagrams are plotted for the case of compensated built-in field, $E_{SI} + E_a = 0$. The thickness $h = 50$ nm, and ion formation energies $\Delta G_1^{00} = \Delta G_2^{00} = 0.2$ eV.

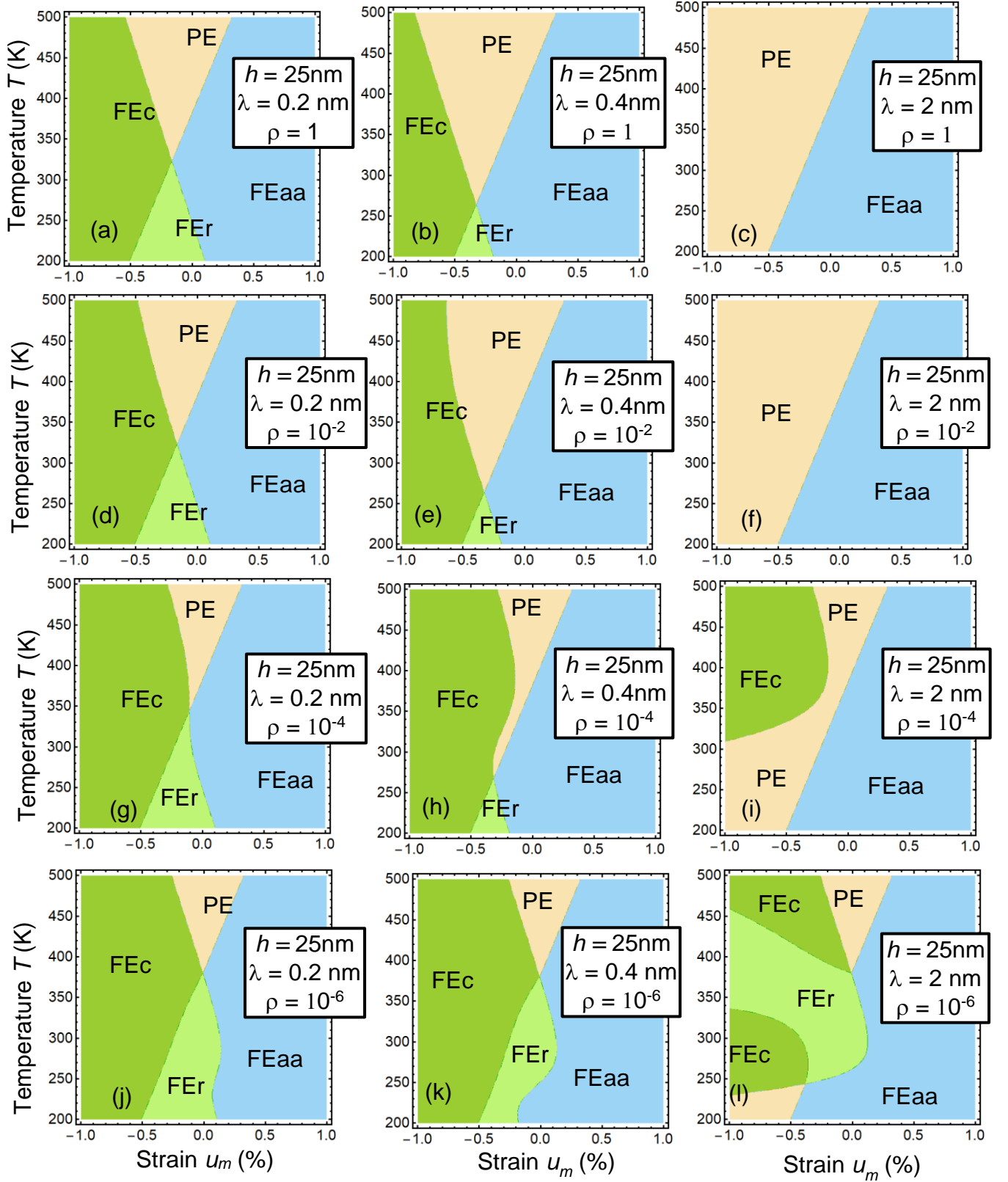

**FIGURE S2.** Typical phase diagrams of a multiaxial FE film in dependence on the temperature $T$ and mismatch strain $u_m$. The values of relative partial oxygen pressure, $\rho = 1, 10^{-2}, 10^{-4}, 10^{-6}$, and gap thickness, $\lambda = 0.2, 0.4, 2$ nm, are listed for each plot. The diagrams are plotted for the case of compensated built-in field, $E_{SI} + E_a = 0$. The thickness $h = 25$ nm, and ion formation energies $\Delta G_1^{00} = \Delta G_2^{00} = 0.2$ eV.

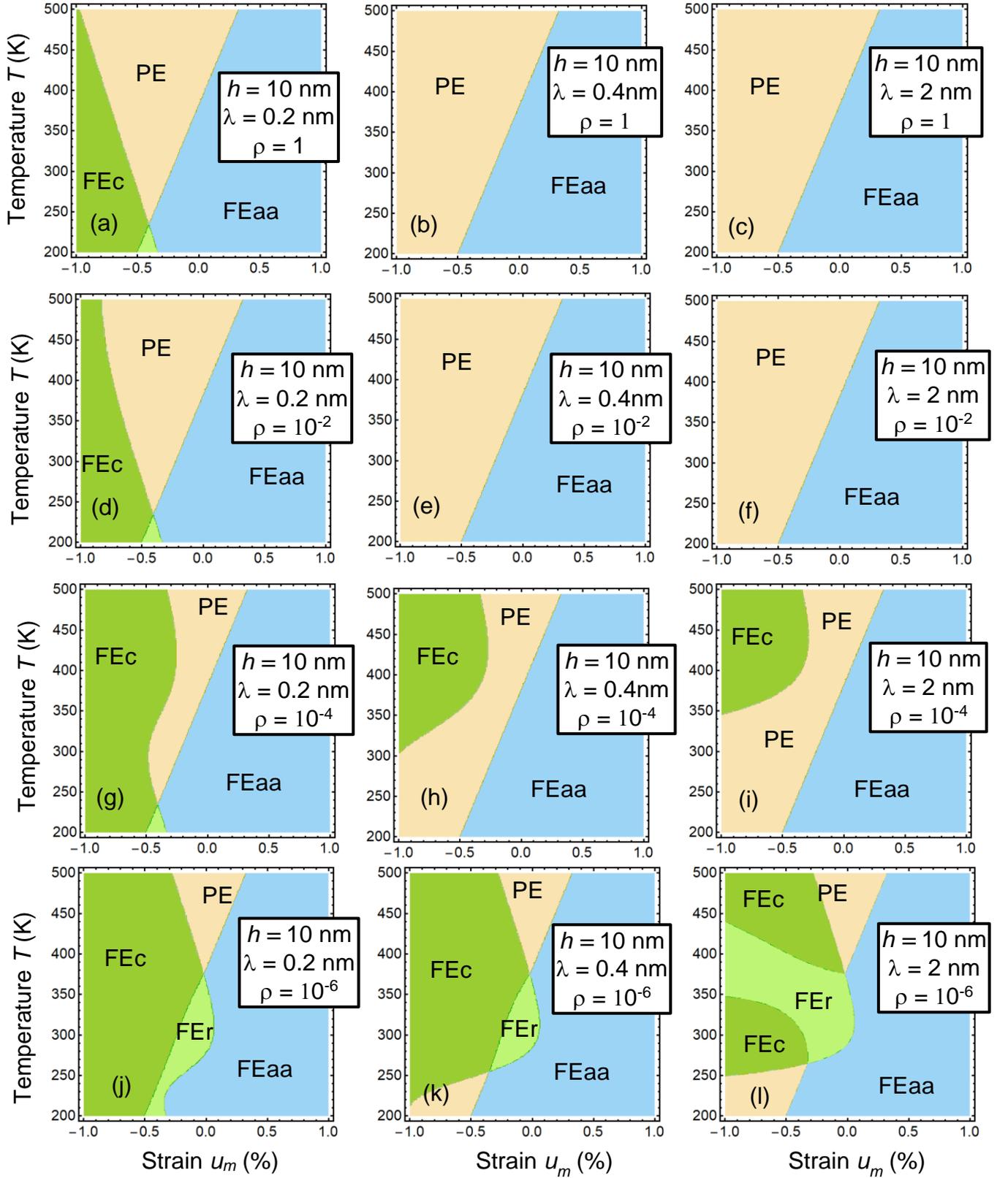

**FIGURE S3.** Typical phase diagrams of an ultra-thin multiaxial FE film in dependence on the temperature $T$ and mismatch strain $u_m$. The values of relative partial oxygen pressure, $\rho = 1, 10^{-2}, 10^{-4}, 10^{-6}$, and gap thickness, $\lambda = 0.2, 0.4, 2$ nm, are listed for each plot. The diagrams are plotted for the case of compensated built-in field, $E_{SI} + E_a = 0$. The thickness $h = 10$ nm, and ion formation energies $\Delta G_1^{00} = \Delta G_2^{00} = 0.2$ eV.

## B. Phase diagrams in coordinates temperature – oxygen pressure

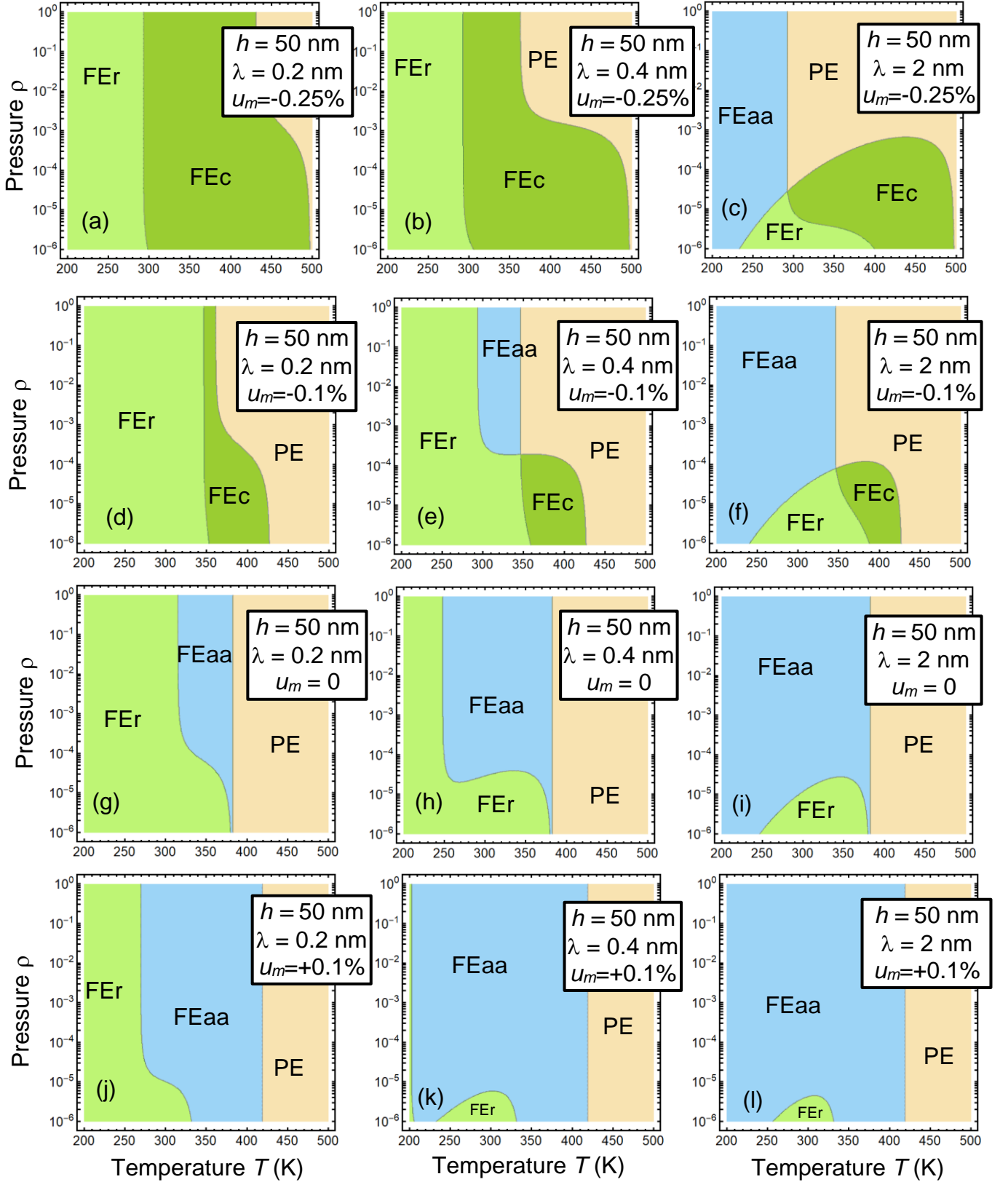

**FIGURE S4.** Typical phase diagrams of an ultra-thin multiaxial FE film in dependence on the temperature $T$ and relative partial oxygen pressure $\rho$. The values of mismatch strain $u_m = -0.25, -0.1, 0, +0.1$, and gap thickness, $\lambda = 0.2, 0.4, 2$ nm, are listed for each plot. The diagrams are plotted for the case of compensated built-in field, $E_{SI} + E_a = 0$. The thickness $h = 50$ nm, and ion formation energies $\Delta G_1^{00} = \Delta G_2^{00} = 0.2$ eV.

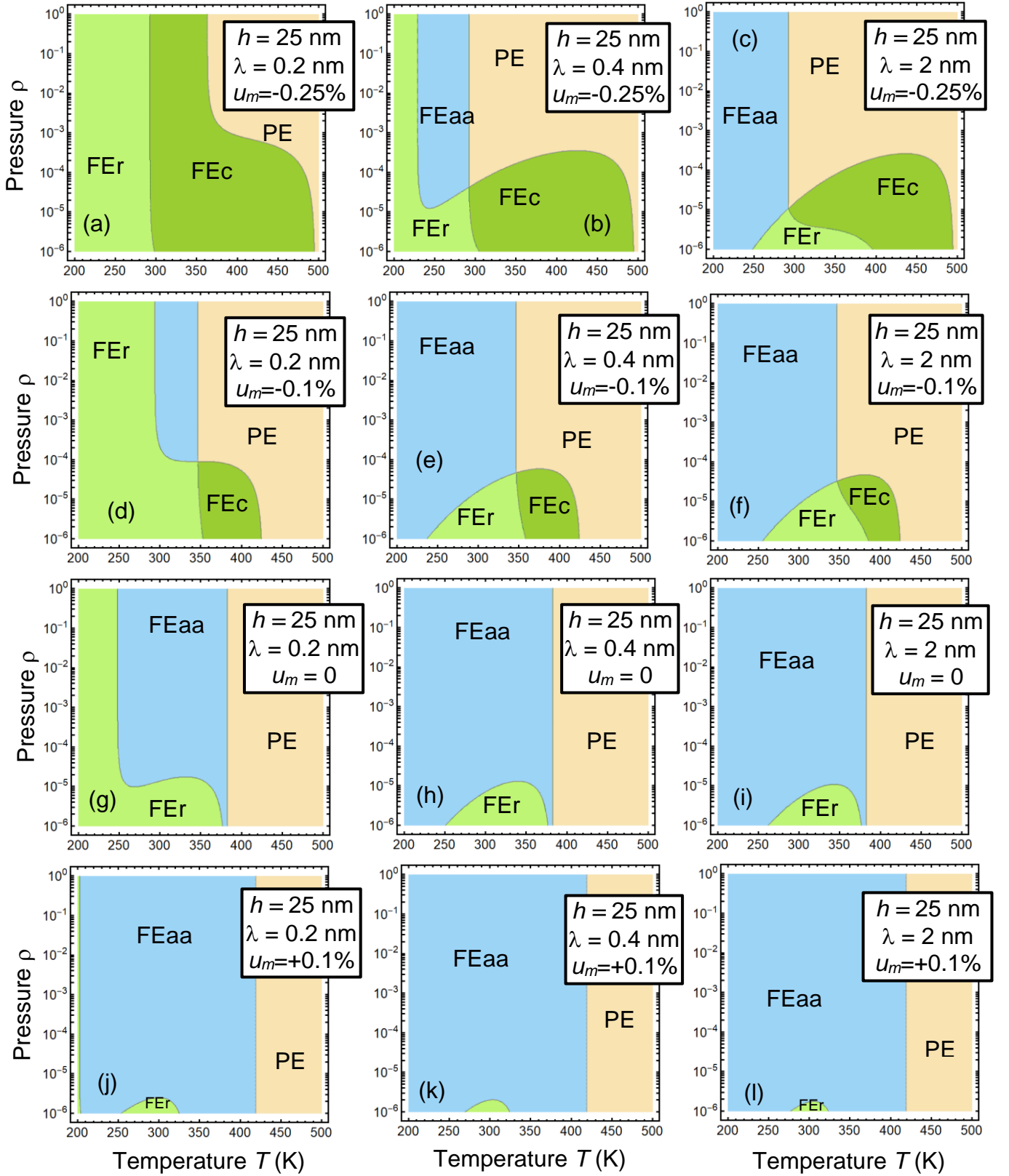

**FIGURE S5.** Typical phase diagrams of an ultra-thin multiaxial FE film in dependence on the temperature $T$ and relative partial oxygen pressure $\rho$. The values of mismatch strain $u_m = -0.25, -0.1, 0, +0.1$, and gap thickness, $\lambda = 0.2, 0.4, 2$ nm, are listed for each plot. The diagrams are plotted for the case of compensated built-in field, $E_{SI} + E_a = 0$. The thickness $h = 25$ nm, and ion formation energies $\Delta G_1^{00} = \Delta G_2^{00} = 0.2$ eV.

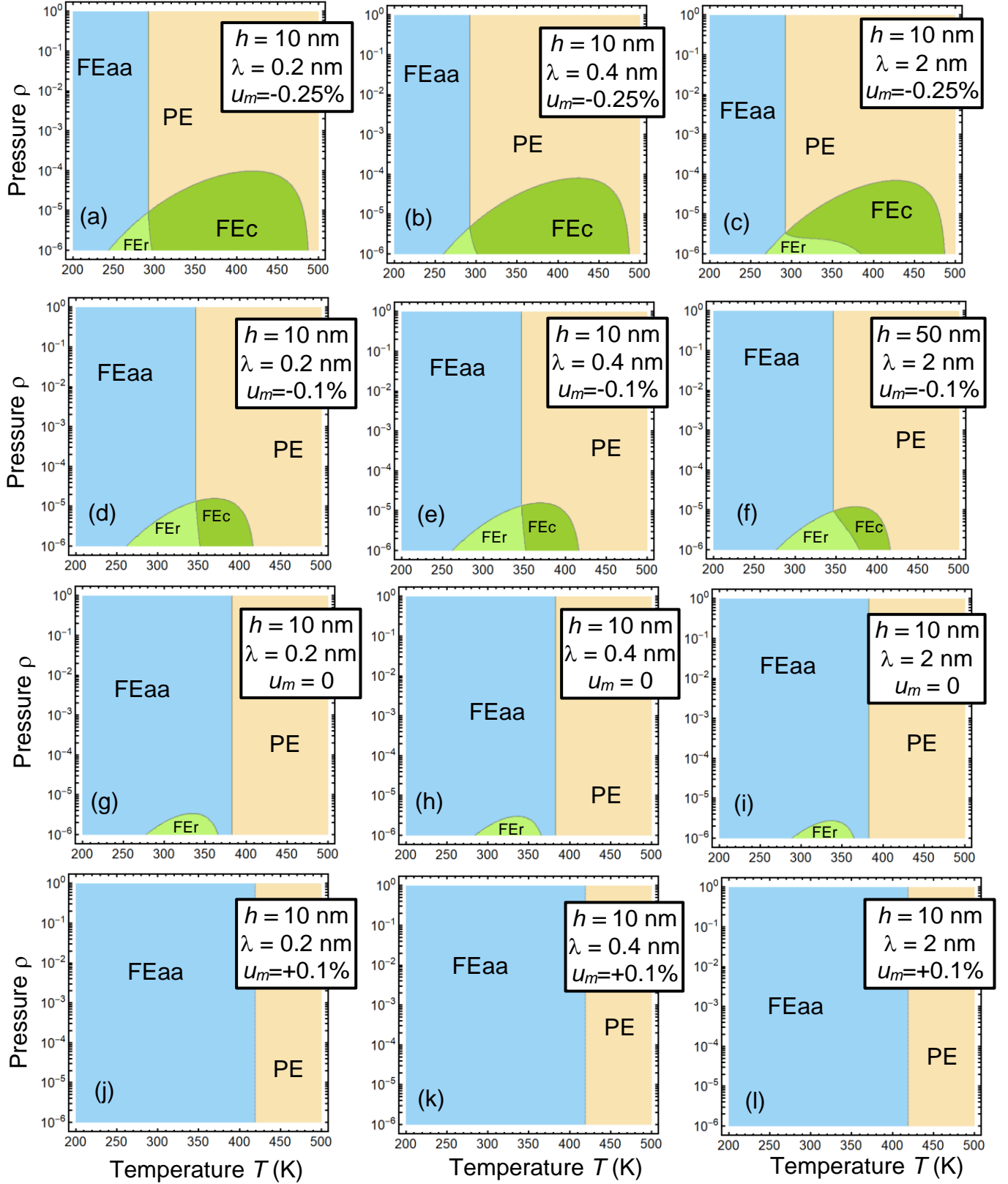

**FIGURE S6.** Typical phase diagrams of an ultra-thin multiaxial FE film in dependence on the temperature $T$ and relative partial oxygen pressure $\rho$. The values of mismatch strain $u_m = -0.25, -0.1, 0, +0.1$, and gap thickness, $\lambda = 0.2, 0.4, 2$ nm, are listed for each plot. The diagrams are plotted for the case of compensated built-in field, $E_{SI} + E_a = 0$. The thickness $h = 10$ nm, and ion formation energies $\Delta G_1^{00} = \Delta G_2^{00} = 0.2$ eV.

## C. Size effect of phase diagrams

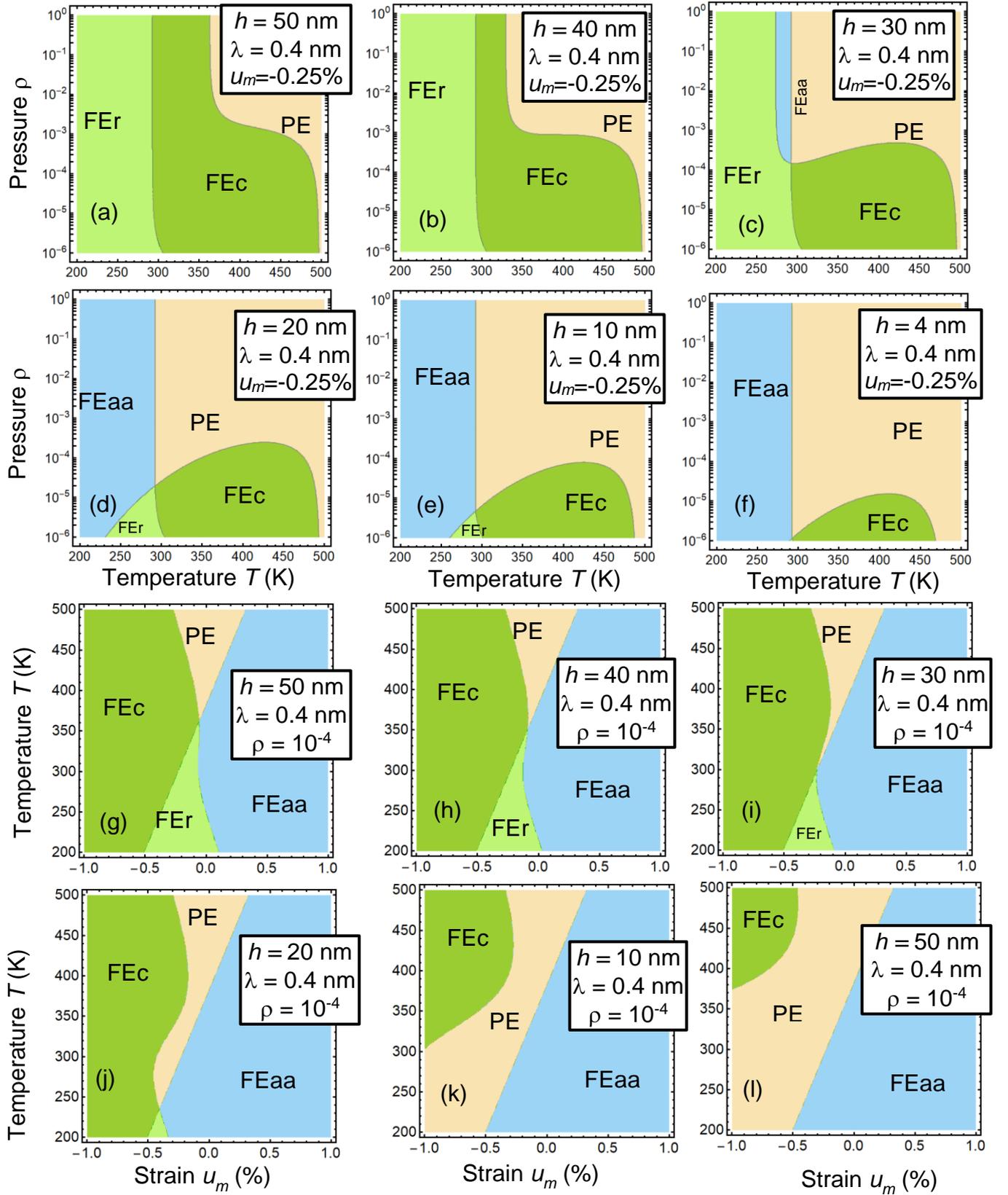

**FIGURE S7.** Typical phase diagrams of an ultra-thin multiaxial FE film in dependence on the temperature $T$, relative partial oxygen pressure $\rho$, and mismatch strain $u_m$. The film thickness h and the gap thickness, $\lambda$ are listed for each plot. The diagrams are plotted for the case of compensated built-in field, $E_{SI} + E_a = 0$,

corresponding to the PE phase and three FE phases, FEc, FEaa and FEr. The ion formation energies $\Delta G_1^{00} = \Delta G_2^{00} =0.2$ eV.

**D. Polarization, dielectric permittivity and piezoelectric coefficients field dependences**

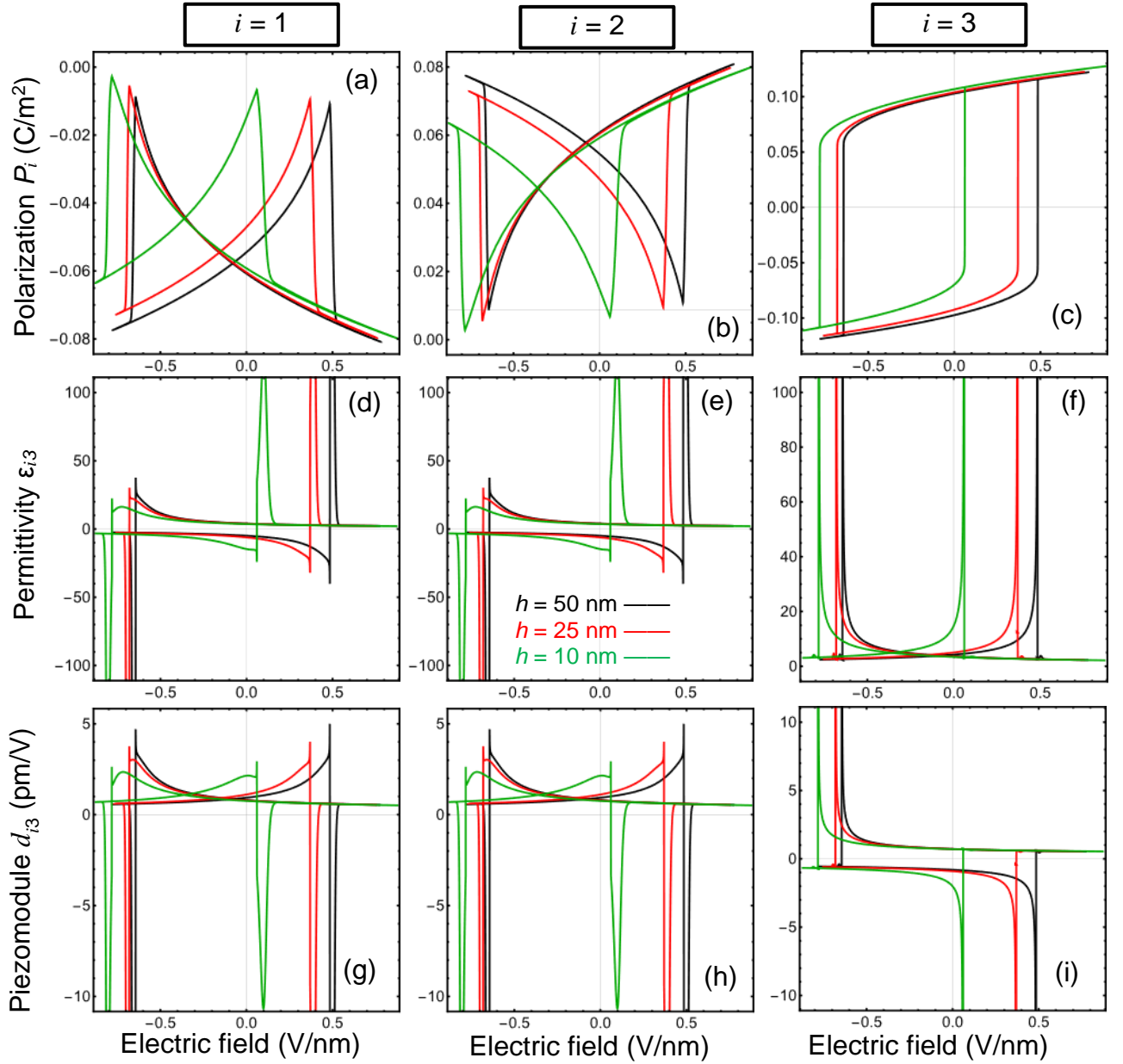

**FIGURE S8** Quasi-static dependences of the polarization components $P_1$ (**a**), $P_2$ (**b**) and $P_3$ (**c**), dielectric permittivity $\varepsilon_{13}$ (**d**), $\varepsilon_{23}$ (**e**) and $\varepsilon_{33}$ (**f**), and piezoelectric coefficients $d_{13}$ (**g**), $d_{23}$ (**h**) and $d_{33}$ (**i**) on external electric field, calculated for the partial oxygen pressure $\rho = 10^{-6}$ and several values of film thickness 50 nm (black curves), 25 nm (red curves), 10 nm (green curves). Other parameters: $\lambda = 2$ nm, $u_m = -0.25\%$ and $T = 350$ K.

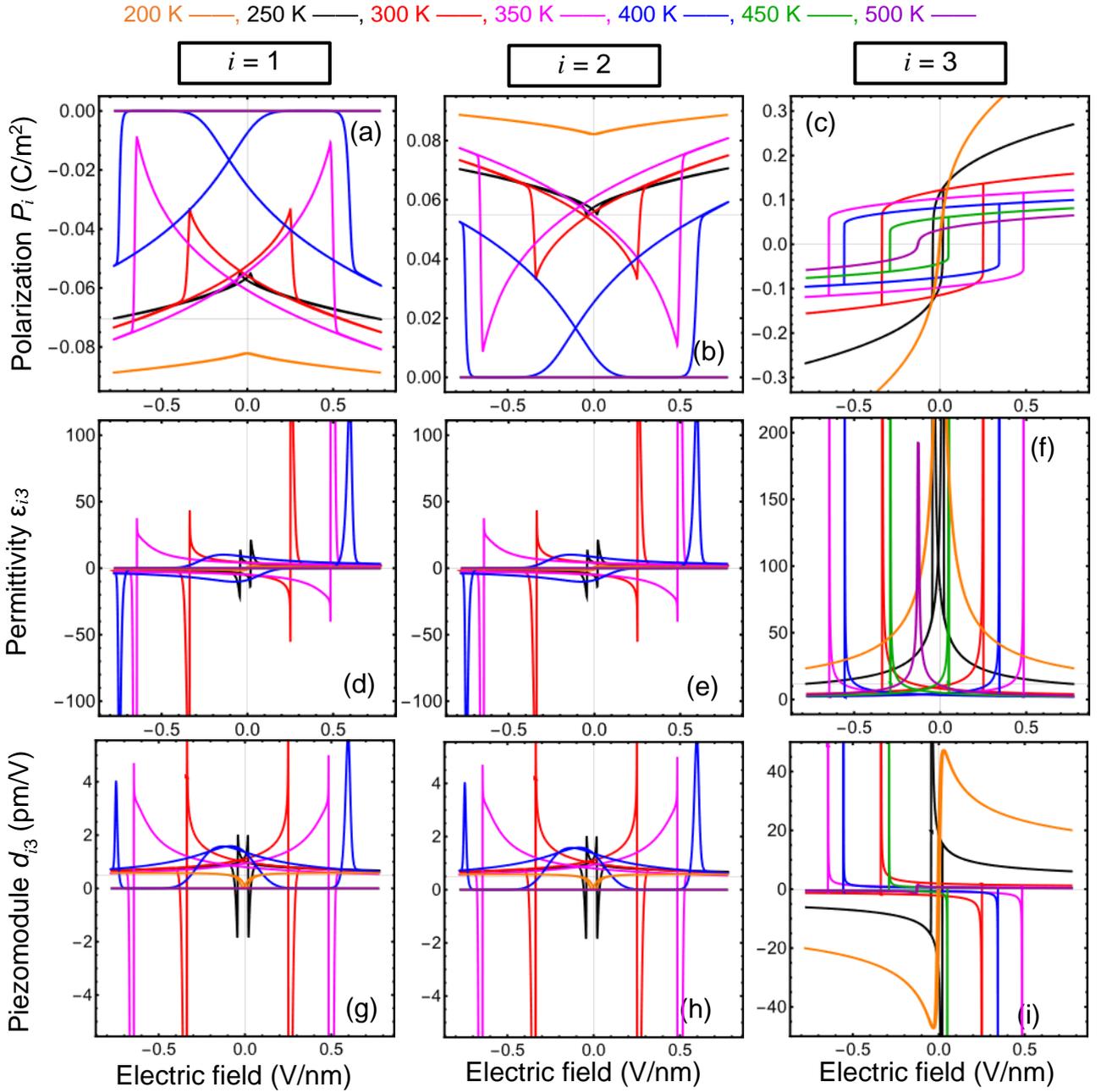

**FIGURE S9.** Quasi-static dependences of the polarization components $P_1$ (a), $P_2$ (b) and $P_3$ (c), dielectric permittivity $\varepsilon_{13}$ (d), $\varepsilon_{23}$ (e) and $\varepsilon_{33}$ (f), and piezoelectric coefficients $d_{13}$ (g), $d_{23}$ (h) and $d_{33}$ (i) on external electric field, calculated for the film thickness 50 nm and several temperatures $T = 200$ K (black curves), 250 K (red curves), 300 K (magenta curves), 350 K (blue curves), 400 K (green curves), 450 K (light blue curves), and 500 K (brown curves). Other parameters: $\lambda = 2$ nm, $u_m = -0.25\%$ and $\rho = 10^{-6}$.

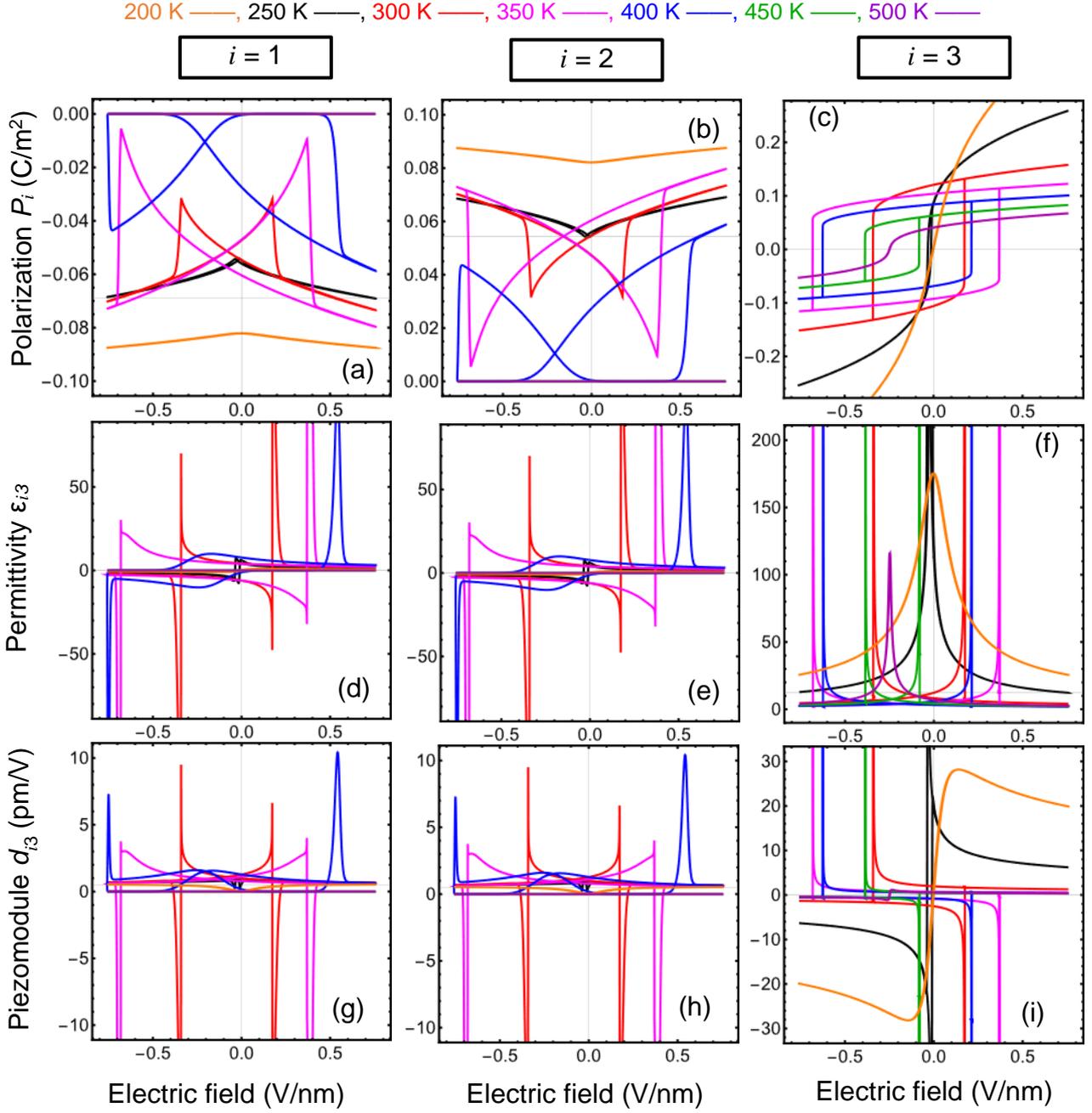

**FIGURE S10.** Quasi-static dependences of the polarization components $P_1$ (**a**), $P_2$ (**b**) and $P_3$ (**c**), dielectric permittivity $\varepsilon_{13}$ (**d**), $\varepsilon_{23}$ (**e**) and $\varepsilon_{33}$ (**f**), and piezoelectric coefficients $d_{13}$ (**g**), $d_{23}$ (**h**) and $d_{33}$ (**i**) on external electric field, calculated for the film thickness 25 nm and several temperatures $T = 200$ K (black curves), 250 K (red curves), 300 K (magenta curves), 350 K (blue curves), 400 K (green curves), 450 K (light blue curves), and 500 K (brown curves). Other parameters: $\lambda = 2$ nm, $u_m = -0.25\%$ and $\rho = 10^{-6}$.

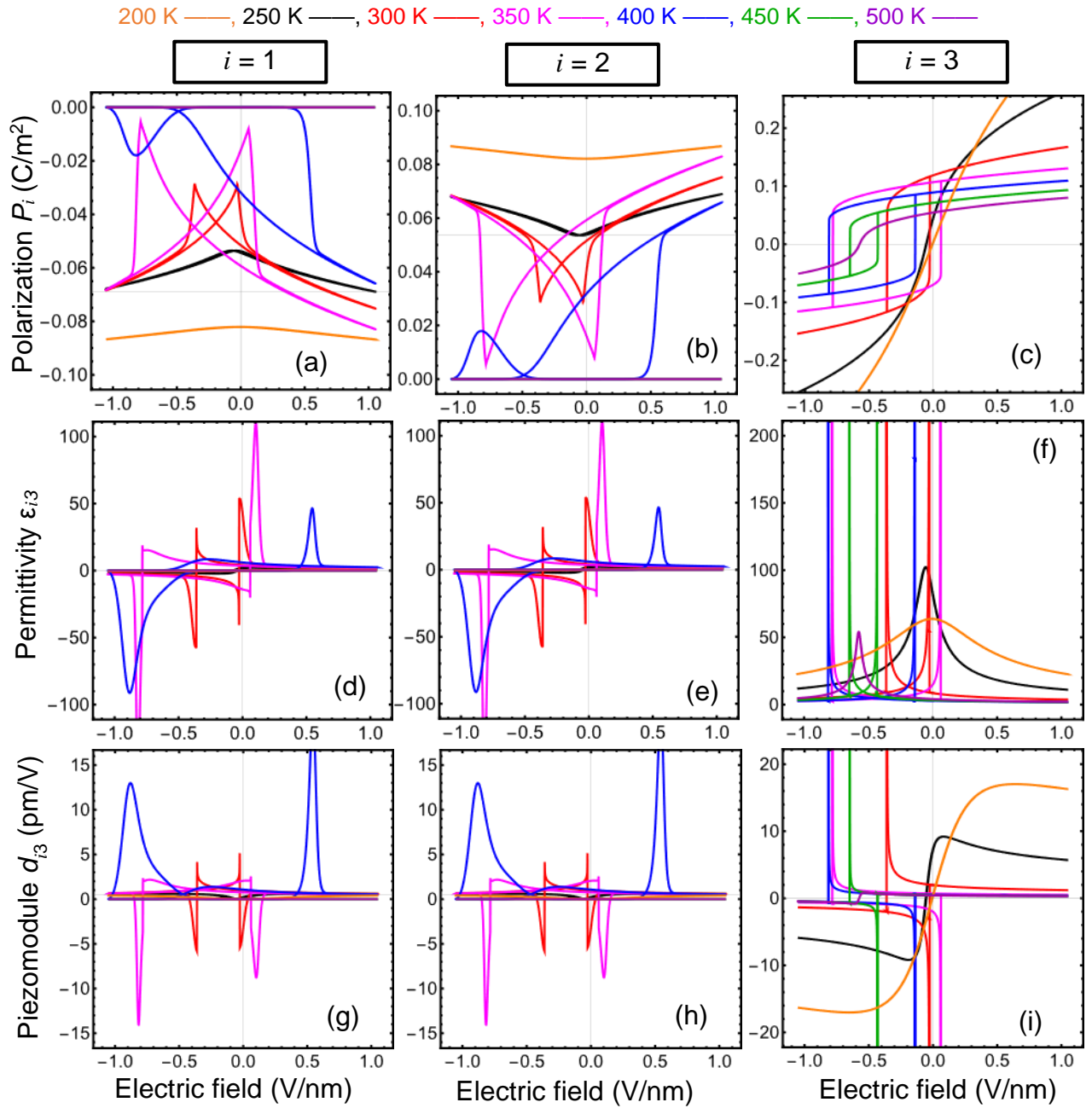

**FIGURE S11.** Quasi-static dependences of the polarization components $P_1$ (a), $P_2$ (b) and $P_3$ (c), dielectric permittivity $\varepsilon_{13}$ (d), $\varepsilon_{23}$ (e) and $\varepsilon_{33}$ (f), and piezoelectric coefficients $d_{13}$ (g), $d_{23}$ (h) and $d_{33}$ (i) on external electric field, calculated for the film thickness 10 nm and several temperatures $T$ = 200 K (black curves), 250 K (red curves), 300 K (magenta curves), 350 K (blue curves), 400 K (green curves), 450 K (light blue curves), and 500 K(brown curves). Other parameters: $\lambda$ =2 nm, $u_m = -0.25\%$ and $\rho = 10^{-6}$.

# APPENDIX D
## Multi-objective Bayesian exploration with Gaussian Process: Formulation and Additional figures

Here, we provide the formulation and description of the functionalities- phase and minimum free energy of a multiaxial FE film in the dependence of the temperature $T(K)$ and mismatch strain $u_m(\%)$. Function 1, $f_1$, defines the phases and function 2, $f_2$, defines the min free energy. Mathematically,

$$f_1 = x + U(0,1), \text{ where } x \in [1, 2, 3, 4, 5, 6] \tag{D.1}$$

$$f_2 = \min(\boldsymbol{f_R^*}) \tag{D.2}$$

Where $x$ is the discrete value between 1 and 6 which maps to the phases in **Table I** as follows: PE – 1, FEa – 2, FEc – 3, FEaa – 4, FEac – 5, FEr – 6. $U(0,1)$ is a random value from uniform distribution, ranges between 0 and 1. $\boldsymbol{f_R^*}$ is the vector of free energy values at all the optimal (local and global) solutions of order parameter vector $\boldsymbol{P} = [P_1, P_2, P_3]$, in minimizing the free energy Eq.(7).

We started with configuring the ground states of both functionalities, using low-sampling (10x10) based exhaustive grid search. Since this is computationally costly, and to analyze further the domain space, we utilize the cheap surrogate Gaussian process tools and develop a multi-objective Bayesian optimization (**MOBO**) architecture. As in many problems like in this paper, there is a general need to focus of different functionalities from expensive evaluations, which is more complex in the field of optimization. Thus, we develop the MOBO architecture, where it first replaces with cheap surrogate models (e.g., Gaussian process) for each chosen functionality and then attempt for rapid exploration/exploitation towards learning interesting domain region through sequential sampling from optimizing a multi-function-based acquisition function. One challenge we have in this problem is the discrete nature of function mapping to phase, which can impact the learning and the prediction accuracy of the GP model, thereby impacting the efficiency of MOBO. In order to make the function smoother, without changing the actual trend, we include a small noise to the function defined by a uniform distribution in the range between 0 and 1 such as, $U(0,1)$. This makes the GP model more efficient, thereby the MOBO is efficient to learn faster (minimize sampling for exploration). For all the analysis in **Figs. 7** and **D1**, the MOBO is started with evaluating 20 randomly selected locations (samples) from the 2-4 LGD model using material parameters of a bulk BaTiO$_3$ from **Table II**, film and interface parameters from **Tables III** and stopped after 230 MOBO guided evaluations from a dense 50x50 map, with a total of 250 expensive evaluations.

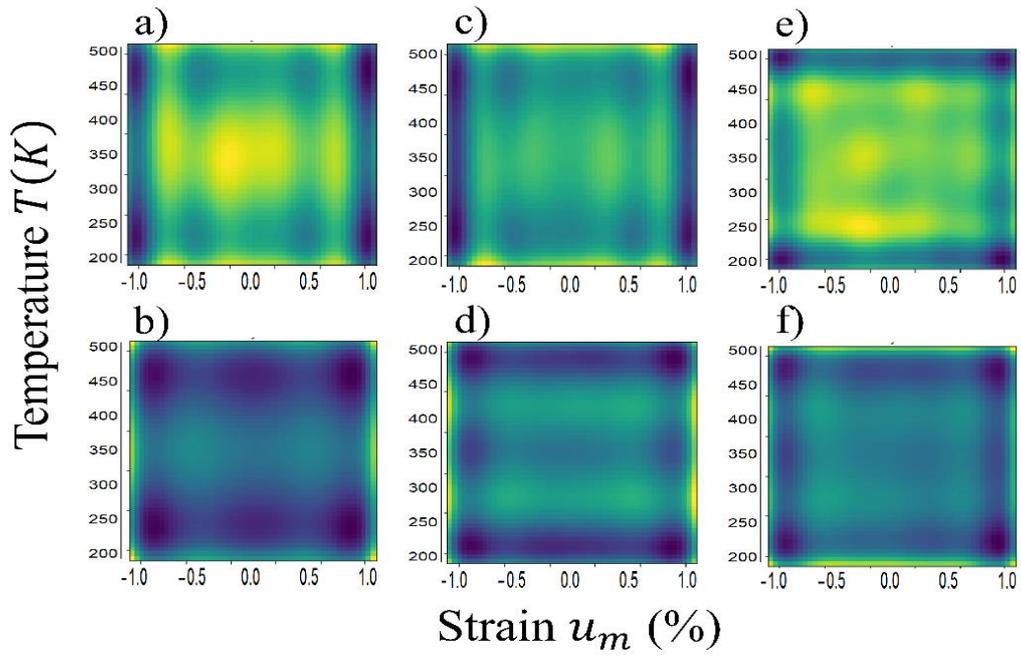

**Figure D1.** Additional figures of **Fig. 7**: Joint exploration of the phase and minimum free energy diagrams of a multiaxial FE film in dependence of the temperature $T$ and mismatch strain $u_m$. Parts **(a, c, e)** are the uncertainty maps of predicted phase diagrams of parameter space as in **Figs. 7 (c, g, k)**. Parts **(b, d, f)** are the uncertainty maps of predicted energy diagrams of parameter space as in **Figs. 7 (d, h, l)**.